\documentclass[twocolumn]{article}
\usepackage{color,soul}
\usepackage{graphicx}
\usepackage[top=2cm, bottom=2cm, left=2cm, right=2cm]{geometry}
\usepackage{authblk}

\newcommand\rev[1]{\textbf{#1}}
\renewcommand\rev[1]{#1}
\newcommand\revd[1]{\textbf{#1}}
\renewcommand\revd[1]{#1}

\begin{document}

	\title{Desorption of neutrals, cations and anions from core-excited amorphous solid water} 

	 \author[1]{R. Dupuy}
	 \author[1]{G. F\'{e}raud}
	 \author[1]{M. Bertin}
	 \author[4]{C. Romanzin}
	 \author[1]{L. Philippe}
	 \author[1]{T. Putaud}
	 \author[1]{X. Michaut}
	 \author[2]{R. Cimino}
	 \author[3]{V. Baglin}
	 \author[1]{J.-H. Fillion} 
	 
	 \affil[1]{Sorbonne Universit\'e, Observatoire de Paris, Universit\'e PSL, CNRS, LERMA, F-75005, Paris, France}
	 \affil[2]{Laboratori Nazionali di Frascati (LNF)-INFN I-00044 Frascati}
	 \affil[3]{CERN, CH-1211 Geneva 23, Switzerland}
	 \affil[4]{Laboratoire de Chimie Physique, CNRS, Univ. Paris-Sud, Universit\'e Paris-Saclay, 91405, Orsay, France}

	\date{\vspace{-1.5cm}}
	
	%\maketitle

\twocolumn[
  \begin{@twocolumnfalse}
    \maketitle
    
\begin{abstract}
Core-excitation of water ice releases many different molecules and ions in the gas phase. Studying these desorbed species and the underlying mechanisms can provide useful information on the effects of X-ray irradiation in ice. We report a detailed study of the X-ray induced desorption of a number of neutral, cationic and anionic species from amorphous solid water. We discuss the desorption mechanisms, and the relative contributions of Auger and secondary electrons (X-ray induced Electron Stimulated Desorption) and initial excitation (\rev{direct desorption}) as well as the role of photochemistry. Anions are shown to desorb not just through processes linked with secondary electrons but also through direct dissociation of the core-excited molecule. The desorption spectra of oxygen ions (O$^+$, OH$^+$, H$_2$O$^+$, O$^-$, OH$^-$) give a new perspective on their previously reported very low desorption yields for most types of irradiation of water, showing that they mostly originate from the dissociation of photoproducts such as H$_2$O$_2$.   
\end{abstract}
\vspace{0.3cm}
  \end{@twocolumnfalse}
  ]

% insert suggested keywords - APS authors don't need to do this
%\keywords{}

%\maketitle must follow title, authors, abstract, \pacs, and \keywords
%\maketitle

% body of paper here - Use proper section commands
% References should be done using the \cite, \ref, and \label commands
\section{Introduction}

Water ice is of primary importance in numerous domains of science. It is ubiquitous on Earth but also in space, on various bodies of the solar system and on the dust grains of the interstellar medium, where it plays a primordial role in e.g. planet formation in protoplanetary disks\cite{vandishoeck2013,vandishoeck2014c}. The interaction of water ice with energetic radiation such as UV or X-rays interests different fields, such as planetary science\cite{grannas2007} or astrochemistry: water frozen on interstellar dust grains can be released to the gas phase through a process called photodesorption, which has shown its importance in the field over the last decade\cite{dupuy2018,hogerheijde2011,kamp2013,mitchell2013}. Amorphous solid water is also often used as a model for liquid water to which a different variety of experimental techniques can be applied. It can therefore be used to study e.g. radiation effects in liquid water, which is relevant in many fields like biology or nuclear reactors\cite{garrett2005,alizadeh2012}, but also for understanding the structure of condensed water itself. Numerous studies have been dedicated to elucidating the structure of various forms of condensed phase water\cite{nilsson2010,bartels-rausch2012,palmer2018}, especially liquid water, but the latter still eludes complete characterization. X-ray absorption spectroscopy of liquid and different forms of solid water have been a major element in the recent discoveries and controversies surrounding the structure of the different forms of condensed water\cite{nilsson2010}. 

In this context, having a more fundamental understanding of the various processes occurring during X-ray irradiation of water ice is important. One way of getting insights into the dissociation of molecules, photochemistry and energy relaxation pathways is to study under high vacuum conditions thin films of water ice and to investigate what is ejected in the gas phase. This gives information different and complementary to that obtained using techniques that probe the condensed phase. Previous studies of core excitation of water ice by X-rays have mostly investigated the desorption of the H$^+$ ion, which is by far the most abundant ion desorbed. The desorption of other fragments like O$^+$ or OH$^+$ is reported as being particularly low for not only X-ray core excitation\cite{coulman1990} but also other types of irradiations such as electrons\cite{herring-captain2005} or XUV photons\cite{rosenberg1981}. 

Some studies have attempted to derive structural parameters of the ice from the desorption spectrum of H$^+$\cite{rosenberg1983,parent2002}. Another work\cite{coulman1990} looked into the details of the dissociation dynamics by investigating specific features that appear in the H$^+$ desorption spectrum but not in the absorption of the ice, in particular the lowest energy peak of the spectrum, sometimes called the pre-edge, which is shifted and enhanced in the H$^+$ spectrum. This feature was interpreted as pertaining to surface molecules, and it was suggested that an ultrafast dissociation process (i.e. breaking of the O-H bond on the potential energy surface of the core hole state, before its decay) could be occurring. It was later found by Auger electron - ion coincidence (AEPICO)\cite{mase2003} and investigation of the desorption of neutral molecules\cite{romberg2000}, especially neutral H, and kinetic energy distributions\cite{weimar2011}, that no ultrafast dissocation took place, but still a "fast" desorption process was at play, with a significant elongation of the O-H bond during the core hole lifetime. This illustrates the importance of exploring the exact mechanisms of desorption in order to use it to derive information on the investigated system. Indeed, the pre-edge peak is a keystone of the current interpretation of X-ray absorption and desorption spectra of condensed water that are used to derive structural information\cite{nilsson2010}. 

One basic distinction that is usually made, in terms of desorption mechanisms for core excitations, is between desorption mediated by the Auger electron and the subsequent secondary electrons created (which is usually termed X-ray induced Electron-Stimulated Desorption, XESD) from desorption mediated directly by the core excitation decay (as a result of the molecule being left in a highly excited state), \rev{which will be called direct desorption here (and is sometimes referred to as "true" photon-stimulated desorption\cite{coulman1990}). XESD includes both processes induced by the low-energy ($<$ 20 eV) secondary electrons and the secondary ionization events due to scattering of the Auger electron, which can include the creation of highly excited (e.g. doubly ionized) states. Comparison between absorption and desorption spectra can give us information on the contributions of direct desorption and XESD}. Another point that can be discussed in this case is the role that photochemistry plays in the desorption phenomena. \rev{In the case of UV photodesorption from water ice, in addition to several direct desorption mechanisms that have been evidenced in theoretical and experimental works\cite{desimone2013,yabushita2013}, desorption due to chemistry has been shown to play a role\cite{yabushita2013}.} 
 
Here, we investigated the spectrally-resolved desorption from core-excited water ice for not only H$^+$ but also most of the other species that could be detected in our set-up. The objective is to make a survey of the desorption of different species upon core-excitation. Desorption of neutral species, protonated clusters, and cation fragments other than H$^+$ are discussed and compared with results from other types of irradiation (electrons, UV photons and energetic ions) when such results exist in the literature. In each case, we attempt to discuss desorption mechanisms when possible and whether we can distinguish the contributions of XESD and direct desorption. The desorption spectra of oxygen-bearing ions bring a new perspective to their low desorption yields. We also investigate the desorption of anions, which had not been hitherto reported for any type of irradiation of water ice to our knowledge, to the exception of low energy electrons, where the specific dissociative electron attachment (DEA) process can occur, \rev{along with dipolar dissociation - which we show to not be the only processes at play here.}         

\section{Methods}

\subsection{Experimental set-up}

Experiments were performed in the SPICES 2 set-up. Some aspects of the experiments have already been detailed elsewhere\cite{dupuy2018}. Briefly, the set-up is an ultra-high vacuum chamber equipped with a closed-cycle helium cryostat, reaching a base temperature of 15 K at the sample holder and a base pressure of $\sim 1 \times 10^{-10}$ mbar at 15 K. The substrate used was a technical copper surface (polycrystalline OFHC copper), electrically insulated from the holder by a kapton foil. This allows to measure the sample current generated by X-ray absorption, when Auger and secondary electrons escape the surface of the ice. This total electron yield (TEY) is considered to represent the absorption of the ice. Water (liquid chromatography standard, Fluka, purified by freeze-pump-thaw cycles) vapour was injected through a dosing tube to grow a $\sim$ 100 monolayers (ML) thick ice on the substrate at 90 K, to yield a compact amorphous solid water\cite{kimmel2001} (c-ASW). The thickness ensures a negligible substrate influence on desorption. Experiments were performed at either 90 K or 15 K. Cooling to 15 K is not expected to change the compact amorphous structure of the ice. 

The set-up was installed on the SEXTANTS beamline of the SOLEIL synchrotron. During irradiation, the photon energy was scanned typically between 520 and 600 eV. The monochromatized beam had a resolution of 150 meV and a flux of 1.4 $\times$ 10$^{13}$ photon.s$^{-1}$ for the experiments on neutral molecules, and a resolution of 80 meV and flux of 2.8 $\times$ 10$^{12}$ photon.s$^{-1}$ for the experiments on ions, constant over the whole range except around 535 eV where a dip was present. The spot at the surface was approximately 0.1 cm$^2$ large. The beam was incident at 47$^{\circ}$ relative to the surface normal, and the polarization was set to horizontal so that at the surface the light had a half out-of-plane and half in-plane components. The absolute energy scale was set so that the pre-edge (thereafter peak 1) of bulk water ice absorption was at 535 eV, which is what is usually done in the literature \cite{parent2002,nilsson2010}. We had not shifted the scale in our previous study \cite{dupuy2018}, but it does not change any conclusions: it is only more convenient to compare the results of different papers. This re-calibration was cross-checked to be coherent with the position of the main resonance peak of solid CO near the O 1s edge, which was studied during these same experimental runs. 

Neutral species desorption was detected using a quadrupole mass spectrometer (Pfeiffer Vacuum). Ion desorption was detected with another quadrupole mass spectrometer (EQS Hiden Analytical), which can detect both positive and negative ions. This QMS is equipped with a 45$^{\circ}$ deflector kinetic energy analyser\cite{hamers1998}. It is important to note that kinetic energy filtering is not optional with such a device: all the spectra that we present are taken at a given kinetic energy, the center of the KE distribution unless stated otherwise. Because we have evidence that kinetic energy filtering can have an effect on the relative intensities of the spectral features for some ions (which will not be discussed further here), KE distributions at different fixed photon energies were measured and integrated to check that the KE-differentiated spectra we present are not distorted compared to the KE-integrated spectra. 

\subsection{Calibration of the photodesorption yields}

We estimated the absolute photodesorption yields of the different species we observed. The derivation of absolute yields for neutral species were presented in detail previously\cite{dupuy2018}. For deriving the yields of ions, several steps are necessary. Since our QMS is equipped with a kinetic energy filter, we first integrated the kinetic energy distribution of each ion. The result is corrected by the photon flux at the relevant energy. Then in order to compare different ions between themselves, we estimated the relative detection efficiencies. The assumption is that the relative detection efficiency only depends on the mass of the ion. Estimations made with different gases with known cracking patterns and compared using a calibrated pressure gauge allowed to derive an apparatus function for the QMS, which roughly follows a (m/z)$^{-0.5}$ power law (except for the very light H$^+$ and H$_2^+$ ions, for which separate estimations of the apparatus function were made). The last step is to calibrate the absolute desorption yield of at least one ion. The solution we adopted for now was to scale our data with published desorption yield values of a given system. In this case we irradiated CO ice and scaled the data we obtained for the C$^+$ ion on the absolute values of Rosenberg et al.\cite{rosenberg1985}. This yielded a detection efficiency for this ion of $\sim$2\% with our QMS.

We are confident with the values given for neutral species, within the uncertainty of $\pm$40\% that we estimated\cite{dupuy2018}. The absolute values for cations (and anions) rely on the validity of the measurements of Rosenberg et al. The relative comparison of cations, however, is valid within the $\pm$40\% error of the apparatus function calibration. The inter-comparison of cations and anions, on the other hand, relies on the assumption that the detection efficiency still only depends on the mass, which given that other parameters of the QMS change may not be true. The calibration of anions should therefore be considered as an order of magnitude estimate. 

\section{Results and discussion}

\subsection{Ice absorption spectroscopy and structure}

\begin{figure}
    \includegraphics[trim={0cm 0cm 0cm 0cm},clip,width=\linewidth]{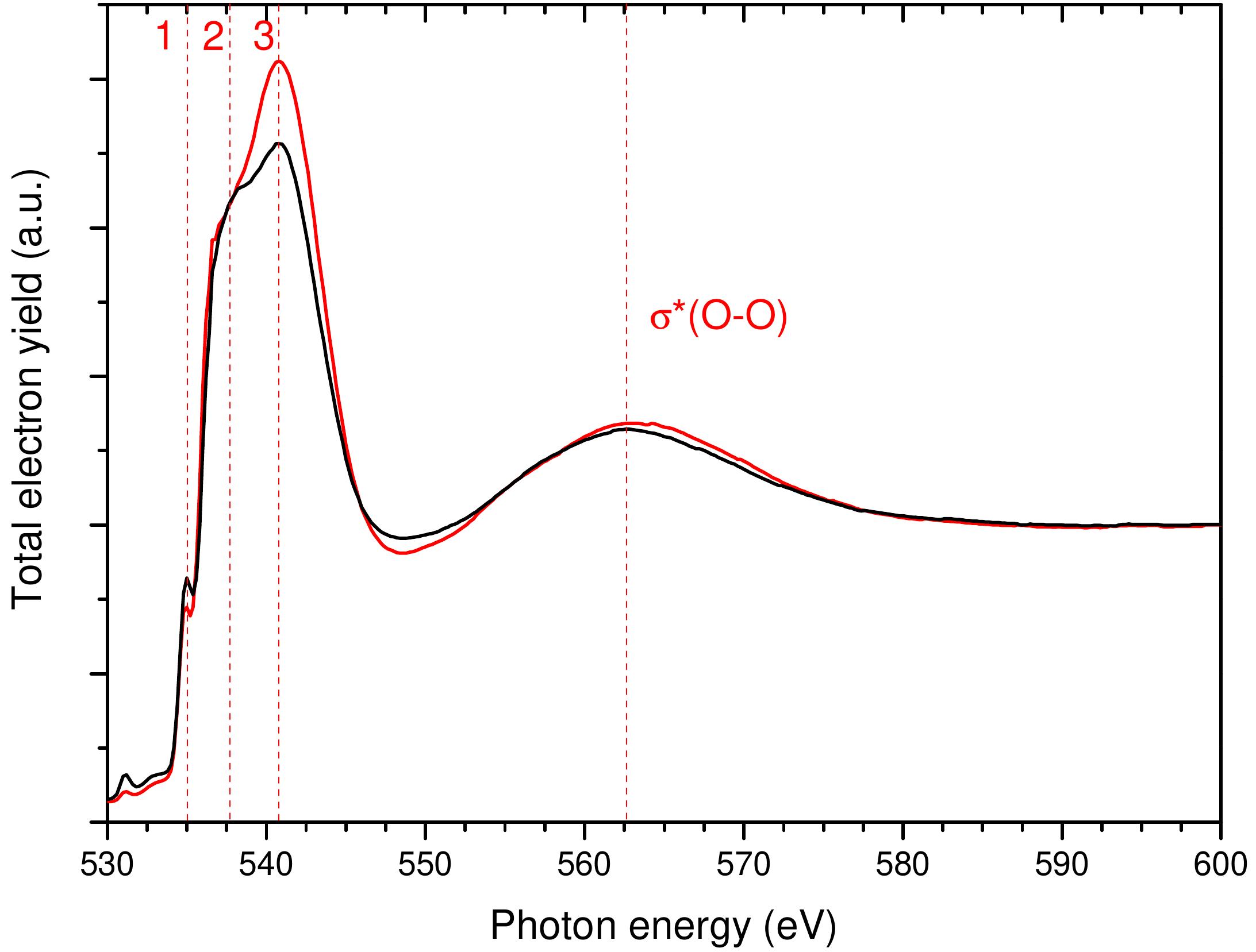}
    \caption{Total electron yield (TEY) as a function of photon energy at 15 K (black trace) and 90 K (red trace) from c-ASW grown at 90 K. The curves have been normalized at 600 eV. The water features are indicated by vertical dashed red lines (see the text for attributions).} 
    \label{TEYs}
\end{figure}

As explained in section II, the measured total electron yield (TEY), i.e. the number of electrons escaping the surface of the ice per incident photon, will be assimilated to the absorption of the ice. The dominant contribution to the TEY are the secondary electrons created by the Auger cascade. The probing depth of the technique, corresponding roughly to the radius of the electron cascade, is about 100 \AA~($\sim$ 30 ML)\cite{timneanu2004}. The contributions of the bulk of the ice are thus presumably dominant, although the surface roughness of amorphous ice may increase the surface contributions and make the estimation of probing depth more complicated. As stated before, contribution of the substrate to the TEY can be neglected given the thickness of the ice (100 ML). The total electron yield curves obtained in our experimental conditions at 15 and 90 K are presented in fig. \ref{TEYs}. 

Interpretation of the TEY spectra as well as the photodesorption spectra that will be presented later requires first a recall of the electronic structure of water and the extensive literature existing on the interpretation of X-ray absorption spectroscopy of condensed water. The free water molecule has three outer valence molecular orbitals 1b$_1$, 3a$_1$ and 1b$_2$ and an inner valence orbital 2a$_1$, as well as a core 1a$_1$ orbital which is the almost unperturbed oxygen 1s orbital\cite{jorgensen1973}. The 1b$_2$ and 2a$_1$ orbitals mostly constitute the O-H bonds, while the 1b$_1$ and 3a$_1$ make up the oxygen lone pairs. There are two empty molecular orbitals below the ionisation threshold of water, the 4a$_1$ and 2b$_2$ orbitals, respectively strongly and weakly antibonding. Above these orbitals lie higher Rydberg states. The absorption spectrum of gas phase water around the O 1s region exhibits four successive peaks before a ionization continuum\cite{wight1974}, which are thus simply attributed to promotion of the 1s electron to, in order, the 4a$_1$, 2b$_2$, and Rydberg orbitals. 

Upon condensation, according to absorption and photoelectron spectroscopy experiments\cite{kobayashi1983}, the valence electronic structure of water is not heavily modified. Shifts and broadening are observed for the filled orbitals. The 4a$_1$ empty orbital is below the conduction band and in condensed phase language would correspond to a Frenkel exciton. The 2b$_2$ orbital on the other hand overlaps with the ionisation threshold and constitutes a conduction band.

In condensed phase, all spectra of water in the core excitation region display three features with varying relative intensities, which we will simply call peak 1, 2 and 3 here for reasons explained below. These are seen on our TEY spectra (fig. \ref{TEYs}) at 535, 537.5 and 541 eV respectively. The peak at 563 eV is the first EXAFS oscillation, which we label $\sigma$*(O-O) to signify its inter-molecular resonance character. Most early works on water core excitation\cite{rosenberg1983,tronc2001,parent2002} attributed the first three features according to the gas phase features, with the first peak being inherited from the 4a$_1$ free molecule orbital, the second from the 2b$_2$ orbital, and the third corresponding to Rydberg orbitals that are heavily modified in condensed phase because of their spatial extension. This is in line with more recent work\cite{tse2008} interpreting the first peak as a localized state of core-excitonic nature, therefore little affected by hydrogen bonding and inheriting from the free molecule 4a$_1$ orbital, while the third one corresponds to a state delocalized along the H-bond network, bearing no relation with the free molecule, but coherent with the solid state notion of a conduction band.

Another interpretation of the spectra has also been proposed. As mentioned in the introduction, X-ray absorption spectroscopy and related techniques have been used to fuel the debate regarding the structure of liquid water and other forms of condensed water. The interpretation takes into account the fact that in condensed phase, all water molecules are not necessarily equivalent. Even in the crystalline phases, the picture of a non-distorted, tetrahedrally bonded molecule does not hold due to the presence of the surface, various defects, grain boundaries and admixtures of amorphous ice in even the best crystalline samples. Calculations have shown the X-ray spectrum of water to be strongly dependent on the bonding of the molecule\cite{cavalleri2002}. In particular, single H-bond donor molecules (i.e. with a free hydrogen) exhibit different features from fully coordinated molecules. These calculations in addition to considerable amounts of experimental data on various forms of condensed water (see refs \cite{nilsson2010,sellberg2014} and references therein) lend credit to the following interpretation: peak 1 and 2 of the spectra, called pre- and main-edge in these references, gain intensity from weakly coordinated, and in particular single donor (SD) species, while peak 3 (called the post-edge) gains intensity from fully coordinated species. The key difference between this interpretation and the previous one is whether there can be a localized excitation contributing to the pre-edge feature for fully coordinated species in the ice. The transposition of gas-phase peak attributions to the condensed phase is therefore not so clear, which is why we adopted the labels peak 1, 2 and 3 for this paper.

We will use the second theory as a framework of interpretation of our data. The difference of our TEY spectra at 15 and 90 K can thus be interpreted as follow. We see that at 90 K, the post-edge (peak 3) is more intense and sharper than at 15 K, while the pre-edge (peak 1) is slightly decreased. We can conclude that at 90 K, more fully coordinated species with less distorted H-bond are probed. The ice is grown at 90 K, which yields a compact amorphous structure, and we do not expect cooling it from 90 to 15 K to change that structure. However we are never probing a pristine ice: the high photon flux used modifies the ice, as expected for any type of irradiation\cite{dartois2015}. The irradiation should create defects and trapped species that change the local structure, affecting the hydrogen-bond network. The differences between the two temperatures therefore stem from the fact that at 90 K, the ice more easily regenerates itself, as defects and trapped species can diffuse and molecules can more easily rearrange themselves. This is why more fully-coordinated species with sharper features (peak 3) are observed at 90 K.    

\begin{figure}
    \includegraphics[trim={0cm 0cm 0cm 0cm},clip,width=\linewidth]{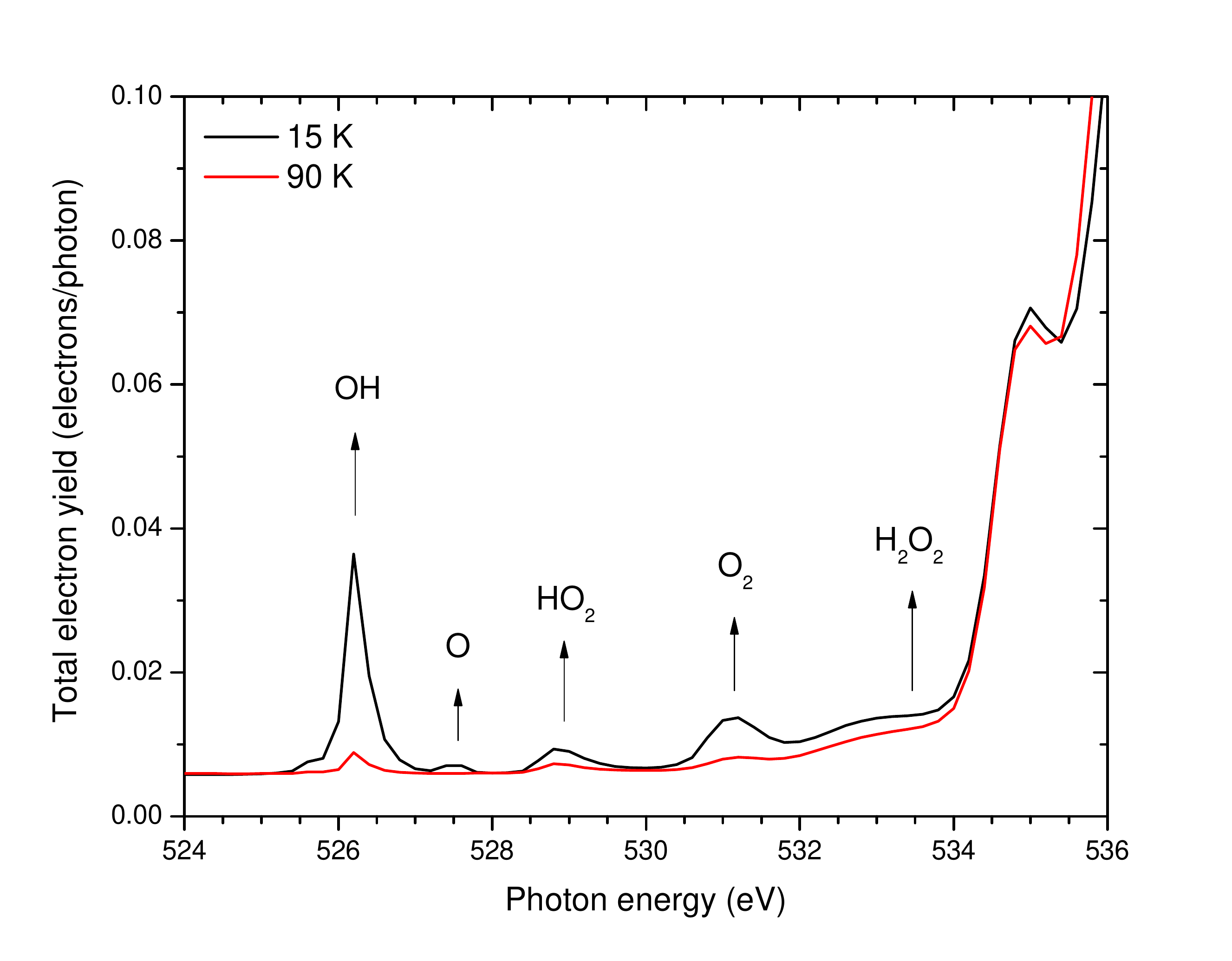}
    \caption{Total electron yield (TEY) as a function of photon energy in the pre-edge region at 15 K (black trace) and 90 K (red trace) from a compact ASW ice grown at 90 K. Attribution of the peaks following Laffon et al.\cite{laffon2006} are indicated.}
    \label{pre_edge_zoom}
\end{figure}

\begin{table*}
      \caption[]{Photodesorption yields (molecules/photon) at 550 eV for c-ASW, at either 15 K or 90 K}
         \label{table_yields}
         \begin{tabular}{c c c c c c c c c}
                    & 90 K & 15 K & & 90 K & 15 K & & 90 K & 15 K \\
             H$_2$O & $3.8 \times 10^{-3}$ & $3.4 \times 10^{-3}$ & H$^+$             & 
$1 \times 10^{-4}$   & $4.3 \times 10^{-5}$ & H$^-$   & $1.3 \times 10^{-5}$ & $6.5 \times 10^{-6}$ \\
             H$_2$  & $8.9 \times 10^{-3}$ & $5.3 \times 10^{-3}$ & H$_2^+$           &
$5 \times 10^{-7}$   & $1.6 \times 10^{-7}$ & H$_2^-$ & $3 \times 10^{-10}$  & 						\\
             O$_2$  & $6.3 \times 10^{-3}$ & $4 \times 10^{-4}$   & H$_3^+$           &
$2.5 \times 10^{-9}$ & 						& O$^-$   & $1.3 \times 10^{-7}$ & $2.3 \times 10^{-8}$ \\
             OH     & $< 1 \times 10^{-3}$ & $< 1 \times 10^{-3}$ & O$^+$             &
$3.6 \times 10^{-8}$ & $2.8 \times 10^{-8}$ & OH$^-$  & $2.2 \times 10^{-8}$ & $6.4 \times 10^{-9}$ \\ 
             		& 					   & 					  & OH$^+$            &
$6 \times 10^{-9}$   & $4.6 \times 10^{-9}$ & H$_2$O$^-$ & $3 \times 10^{-10}$ & 					\\
             		& 					   & 					  & H$_2$O$^+$        &
$3 \times 10^{-9}$   & $3.7 \times 10^{-9}$ & O$_2^-$ & $8 \times 10^{-10}$  & 						\\ 
            		& 					   & 					  & H$_3$O$^+$        &
$1.2 \times 10^{-8}$ & $1.6 \times 10^{-8}$ & 		  & 		 			 & 						\\ 
             		& 					   & 					  & O$_2^+$           &
$2.9 \times 10^{-8}$ & 					    & 		  & 					 & 						\\ 
             		& 					   & 					  & (H$_2$O)$_3$H$^+$ &
$1.9 \times 10^{-8}$ & $2.4 \times 10^{-8}$ &		  & 					 & 						\\    
         \end{tabular}
   \end{table*}

\subsection{X-ray induced fragmentation and chemistry}

Core excitation of a water molecule leads to Auger decay, which (i) leaves the molecule that absorbed the photon in a highly excited state, and (ii) releases an Auger electron. The highly excited states of a molecule that underwent Auger decay are typically doubly ionized (2h: two holes in valence orbitals) states after core ionization and singly ionized excited states (2h1e: two holes in valence orbitals and one electron in a previously empty orbital) after core excitation. These states lead to complex fragmentation patterns. The Auger electron, here with an energy of about 500 eV, will scatter in the ice creating secondary excitations and ionisations - according to simulations, about 25 in the case of water ice\cite{timneanu2004}. Although an energetic electron can create highly excited states in molecules as well, most of these secondary events will create "simple" states, i.e. singly ionised or excited molecules. Part of the focus of this paper, as mentioned in the introduction, will be to distinguish between desorption triggered by the Auger electron cascade (XESD) from direct desorption mediated by the core excitation decay. 

The chemistry induced by X-rays, but also by energetic ions or electrons, sometimes termed radiolysis, is mostly driven by the cascade of secondary electrons and secondary excitations. The resulting outcome is qualitatively similar to photolysis chemistry, i.e. chemistry driven by UV photons (see Yabushita et al.\cite{yabushita2013} for a recent review of water photolysis). The driver of water chemistry is the dissociation of excited H$_2$O$^*$ in OH + H, which is the main dissociation pathway, although direct production of O radicals is also possible. The radicals then react to form further products. In our experiment, some of the main elements of the water chemical network can be identified in situ in the X-ray absorption spectrum of the ice. Figure \ref{pre_edge_zoom} shows the TEY between 524 and 536 eV at either 15 or 90 K. The observed peaks can be attributed to various species other than H$_2$O created by X-ray irradiation in the ice: we see the O, OH and HO$_2$ radicals as well as O$_2$ and H$_2$O$_2$. The intensity of these peaks (at a given temperature) do not change upon further irradiation of the ice, showing that we are probing a steady state. In this steady state, Laffon et al.\cite{laffon2006} estimated from the peak intensities that these chemical products amount to no more than a few percent of the ice. Pure water ice shows a resiliency to irradiation, i.e. most of the photochemistry occurring leads back to H$_2$O as the end product. In photodesorption, two volatile products of the chemistry, H$_2$ and O$_2$ are also observed. The desorption of neutral species is discussed in more details in section C and in a previous publication\cite{dupuy2018}. Other elements that play a role in water photochemistry and cannot be probed here are the important atomic H, the solvated electron, and the presence of defects in the ice.  

Temperature plays a major role in this chemistry, by activating or not the diffusion of radicals. We see in figure \ref{pre_edge_zoom} that at 15 K, the OH radical is much more abundant than at 90 K, a temperature at which it can diffuse easily and react. At 90 K volatile species such as O$_2$ are also more depleted as they can diffuse to the surface and desorb. Part of these species stay trapped in the ice, which has important consequences for astrophysical applications, where the origin of trapped oxygen in ice has been a long-studied \rev{question}\cite{sieger1998,reimann1984} that is not yet fully solved\cite{johnson2011}. \rev{Other temperature effects have been evidenced mainly from the observation of ion yields (H$^+$\cite{sieger1997} and protonated clusters\cite{herring-captain2005}) variations. With increasing temperature there is an increase of hole mobility and of excited state lifetimes, which could also play a role in the evolution of chemistry. This is also linked with structural differences, due to defect migration and reorientation of molecules.}

In addition to the radicals-fuelled chemistry, ionising radiation also opens the possibility of ion chemistry. Contrary to valence excitation, ionization of H$_2$O is not necessarily dissociative\cite{ramaker1983b}. H$_2$O$^+$ ions will therefore be created along with protons and possibly other ions. The released electrons, once thermalized, can be trapped\cite{simpson1998} or mobile\cite{lu2001} and are equivalent to the solvated electron in liquid water. Models of liquid water radiolysis\cite{gervais2006} usually consider that charge or proton transfer processes rapidly make the end product of an ionisation event a hydronium (H$_3$O$^+$) ion and a solvated electron, regardless of the initial product of ionisation. Diffusion of these two species will subsequently lead to recombination and either water reformation or dissociation into radicals, leading back to radical chemistry. However, the fate of the various ions in solid water, where diffusion processes are not necessarily effective, is unclear. Presumably other ionic fragments than H$_3$O$^+$ are also created, although evidence of their presence inside the ice is scarce\cite{johnson2011}. Our detection of various ions in desorption (see table \ref{table_yields}) suggest that they should also exist within the ice, although they may not be abundant or long-lived. For example, the observation of the desorption of protonated clusters (H$_2$O)$_n$H$^+$, detailed in section D, is an indirect evidence of structural rearrangements in the ice of water molecules around a proton, because of ion-dipole interactions. The desorption of anions, detailed in section F and G, shows that they should be present as well. The chemistry induced in water ice by low energy electrons, especially dissociative electron attachment (DEA) and the creation of anionic species, has been studied previously \cite{pan2004}. Desorption can thus provide some insights on the chemistry happening in the ice upon irradiation.

\subsection{Desorption of neutral species}

\begin{figure}
    \includegraphics[trim={0cm 0cm 0cm 0cm},clip,width=0.8\linewidth]{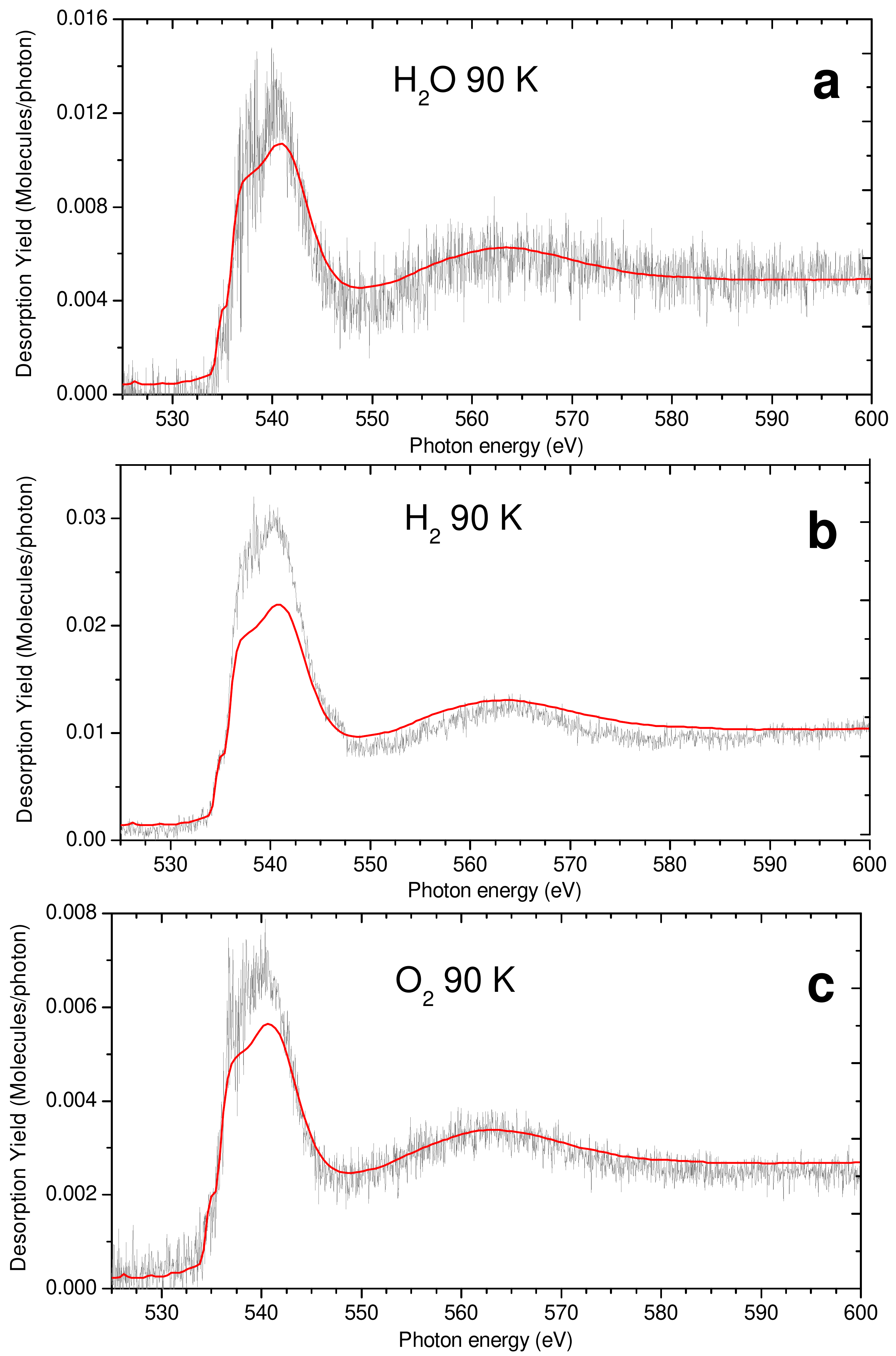}
    \caption{\revd{Photodesorption spectra of neutral molecules from H$_2$O ice at 90 K, with the TEY (red line) shown for comparison, normalized so as to make the different features match best. \textbf{a.} H$_2$O desorption spectrum \textbf{b.} H$_2$ desorption spectrum \textbf{c.} O$_2$ desorption spectrum}}
    \label{neutres}
\end{figure}

\begin{figure}
    \includegraphics[trim={0cm 0cm 0cm 0cm},clip,width=0.8\linewidth]{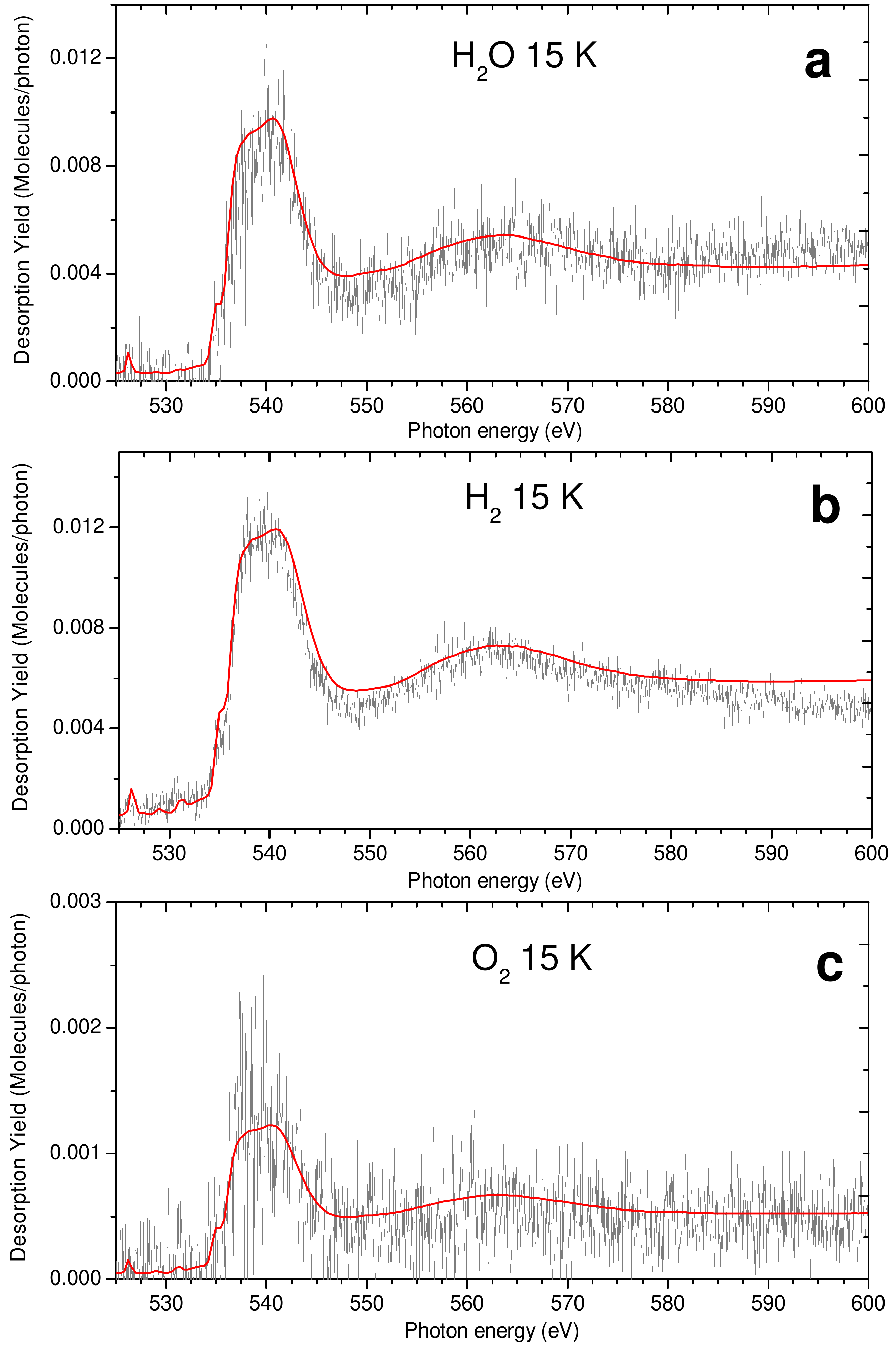}
    \caption{\revd{Photodesorption spectra of neutral molecules from H$_2$O ice at 15 K, with the TEY (red line) shown for comparison, normalized so as to make the different features match best. \textbf{a.} H$_2$O desorption spectrum \textbf{b.} H$_2$ desorption spectrum \textbf{c.} O$_2$ desorption spectrum}}
    \label{neutres2}
\end{figure}

We observed the desorption of three neutral species (table \ref{table_yields}), H$_2$O, O$_2$ and H$_2$. Their photodesorption spectra are given in \revd{figure \ref{neutres} at 90 K and figure \ref{neutres2} at 15 K}. The desorption of neutral species has already been discussed in a previous article\cite{dupuy2018}, therefore the results will only be briefly recalled here along with some new elements. Regarding the distinction between XESD and \rev{direct desorption}, we argued previously that neutral desorption is probably dominated by XESD. The fact that the photodesorption spectra follow the TEY is a first argument, although it is weak: while an XESD process necessarily follows the TEY (electrons created at different photon energies are undistinguishable and will lead to the same effects, therefore these effects should be proportional to the total number of electrons), it is not sufficient: it would also make sense for a \rev{direct desorption} process to follow the absorption spectrum. The argument becomes strong when it is reversed and there is a deviation of the photodesorption spectrum from the TEY, signing unambiguously a \rev{direct desorption} process, as we will see later for ion desorption. 

Another interesting argument is the estimated photodesorption yield per absorbed photon\cite{dupuy2018} (instead of per incident photon) of H$_2$O, which is about 0.2 molecules/photon, which accounting for the uncertainty is close to the yield derived for electron-stimulated desorption\cite{petrik2005}. To better understand the importance of this quantitative argument, we can first point out that XESD dominating for neutral desorption is sensible in the picture we have of how the energy of the initial photon is distributed in the ice. Most of the energy of the initial photon goes into the Auger electron and therefore into the secondary events, which suggests they should dominate desorption. Putting it in numerical terms, since there is one "direct" event for 20-25 secondary events (single ionization + creation of a secondary electron that can cause one valence excitation), if all those events have similar partial cross sections for desorption (within an order of magnitude), then the secondary events will obviously dominate. For this to be true, the secondary events (excitations, ionizations and secondary electrons) need to be energetically sufficient to lead to desorption. Desorption of neutral species has a threshold \rev{around 6.5 eV for electron irradiation}\cite{kimmel1994,kimmel1995}. The secondary electrons are therefore energetically able to desorb neutral species, \rev{presumably through the creation of excitons in the ice and mechanisms similar to those observed in UV irradiation. Another mechanism is electron-ion recombination, so that the many ionization events generated by the scattering of the Auger electron can also lead to desorption.} The only way \rev{direct desorption} could dominate in such a case is if the \rev{direct desorption} process, involving the highly excited initial molecule, is much more efficient than these "simple" excitations to desorb molecules. However, the similarity between the X-ray photodesorption yield per absorbed photon and the electron-stimulated desorption yield does not lend credit to such a possibility. Thus, although a definitive proof is not accessible, we can state with confidence that XESD dominates the desorption of neutral species. 

Desorption of H$_2$ and O$_2$ was observed in irradiation by electrons, ions and UV photons before, and it is not surprising that we see these species during X-ray irradiation as well, considering the chemistry triggered by secondary electrons and excitations is expected to be similar. We could not measure desorption of OH radicals, however our sensitivity limit was rather high ($\sim 1 \times 10^{-3}$ molecule/photon). We did not look for H or O fragments, which were seen desorbing under \rev{low-energy electron\cite{kimmel1995} and} UV irradiation\cite{yabushita2013}, where H is in fact the most abundant desorbing species. Temperature effects on the desorption yields are much more important for the desorption of products of chemistry such as H$_2$ and O$_2$ than for H$_2$O (fig. \ref{neutres} and \ref{neutres2}). Such effects have been studied in detail for other types or irradiation\cite{johnson2011}, and include different contributions. The activation of diffusion of some radicals promotes the creation of these species, and they can also overcome their own diffusion and desorption barriers when the temperature is high enough. 

In the spectra of H$_2$ and O$_2$ desorption at 90 K \revd{(fig. \ref{neutres2})}, a relative increase of the features in the 535-540 eV region relative to the TEY is observed, but it does not seem to have a physical meaning and could instead come from background subtraction issues. 

\subsection{Desorption of protonated clusters}

\begin{figure}
    \includegraphics[trim={0cm 0cm 0cm 0cm},clip,width=\linewidth]{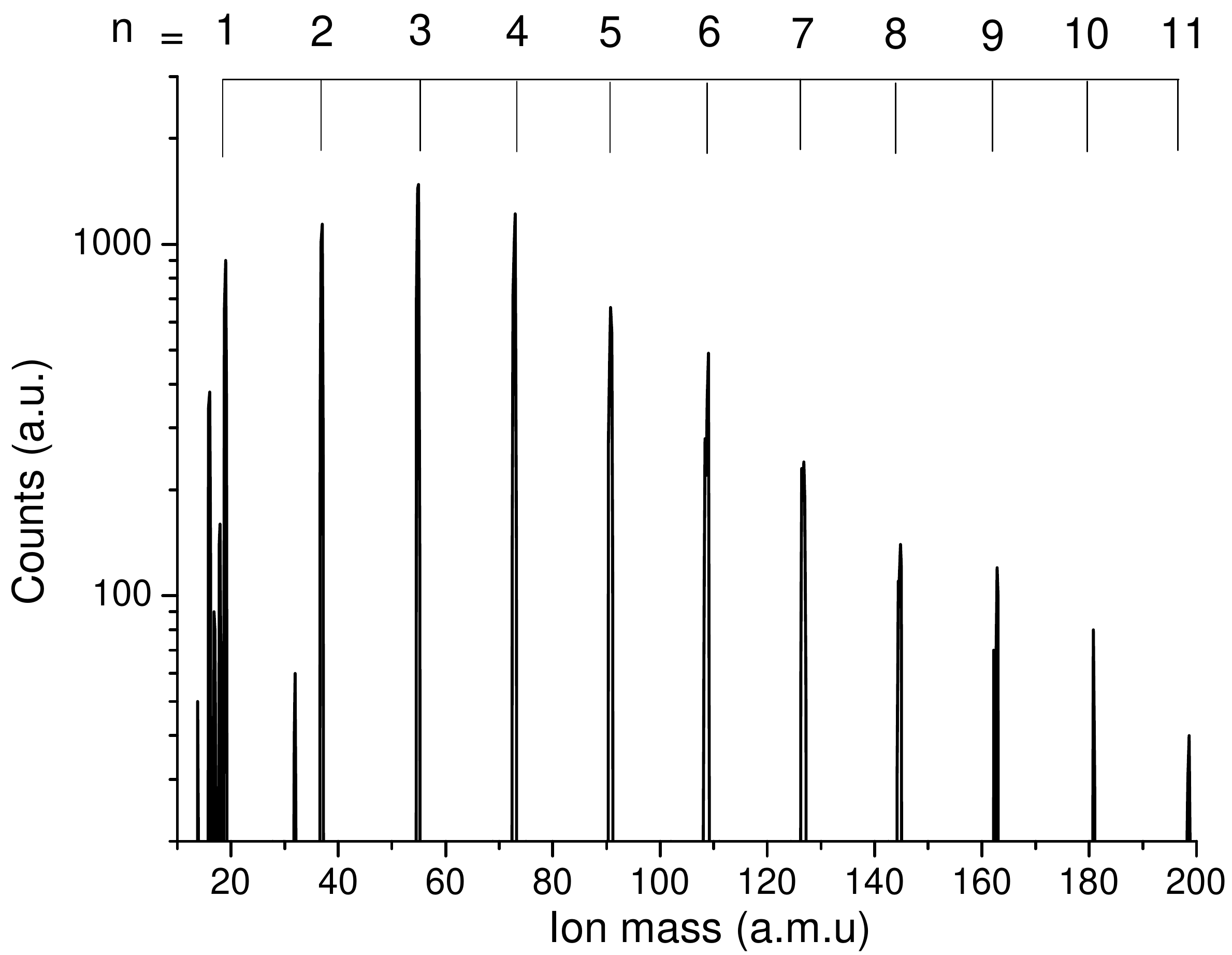}
    \caption{Mass spectrum of the positive ions desorbing at 540 eV and at 15 K from c-ASW. The protonated clusters (H$_2$O)$_n$H$^+$ from n = 1 to 11 are indicated.}
    \label{mspec_clusters}
\end{figure}

The desorption of protonated water clusters (H$_2$O)$_n$H$^+$ was observed up until n = 11 (m/z = 199), which is the limit of our mass spectrometer (m/z = 200). The mass spectrum is shown in figure \ref{mspec_clusters}. Desorption of protonated clusters from water ice by core-excitation was reported before \cite{rocker1990a}. It has also been observed and studied in the case of electron \cite{floyd1972,herring-captain2005} and ion \cite{christiansen1987,martinez2019} irradiation, as well as in field-assisted photon-stimulated desorption\cite{jaenicke1986} or electron and UV photon stimulated desorption from water adsorbed on rare-gas \rev{solids} \cite{souda2002,tachibana2006,grieves2011}.

The formation of protonated clusters in ice exposed to dissociative ionisation is expected, as protons created by ionising dissociation can be stabilized in the form of hydronium (H$_3$O$^+$) ions, onto which a hydration shell can then form due to the strength of the ion-dipole interaction. Based on this microscopic representation, Floyd \& Prince first suggested a model\cite{floyd1972} to explain the observed size distribution of the desorbed clusters. The distribution of the intensities of the clusters, when corrected for the apparatus function of the QMS, peaks at n = 3 in our case, then decreases with increasing cluster size. This distribution reflects the balance of the energy required to break the hydrogen bonds and detach a cluster of size n, and the energy gained from the formation of the ion bonds to the cluster. The former energy increases linearly with n, while the latter decreases quickly beyond addition of the first few molecules. Christiansen et al.\cite{christiansen1987} instead use RRK theory to describe the mass distribution of their clusters. All studies do not present the same maxima for their distributions: while our distribution peaks at n = 3, these of Christiansen et al peak at n = 1, Herring-Captain et al. peak at n = 2, Rocker et al. at n = 2 and Floyd \& Prince at n = 5. These differences could come from different experimental conditions (such as the use of a quadrupole mass spectrometer in our case instead of a TOF mass spectrometer, even though we have corrected our distribution for the apparatus function of the QMS), or from the fact that we use X-rays rather than electrons or ions. In the case of water adsorbed on rare gas films, the maximum varies from n = 1 to n = 3 depending on temperature and coverage \cite{souda2002}, which confirms the role played by the experimental conditions in the maximum of the distribution.  

According to the detailed study of Herring-Captain et al., the mechanism of desorption of cluster ions is distinct from the formation mechanism. The electron energy threshold they observed is above 70 eV \rev{for clusters n $\geq$ 2}, indicating that desorption of clusters requires a complex excitation, such as the formation of a two-hole (2a$_1$)$^{-2}$ state. \rev{Formation occurs by hydration around a hole, similar to what was proposed by Floyd \& Prince. Desorption is instead caused by Coulomb repulsion between the two holes. They argue that the excess energy of the formation of the cluster goes only into their internal modes, and not into translational motion, which purely comes from Coulomb repulsion. One of their main argument is the fact that the measured kinetic energy distribution of the clusters is the same for all values of n. We did observe this similarity of kinetic energy distributions as well (not shown). Emphasis is put in the paper on the fact that Coulomb repulsion occurs for holes located on two neighbouring molecules. After formation of the initial two-hole state there is a transfer of hole: a cluster then forms around an H$_3$O$^+$ and is subsequently ejected by repulsion from the other neighbouring charge. The transfer can occur through e.g. proton transfer but also intermolecular Coulombic decay (ICD), a process whereby the Auger electron is ejected from a neighbouring molecule of the core-ionized one. The importance of ICD in the desorption of clusters was emphasized later by the same group\cite{grieves2011} in a study on the desorption of protonated clusters for slightly different systems (sub-monolayer water adsorbed on graphite or rare gas solids).}

\begin{figure}
    \includegraphics[trim={0cm 0cm 0cm 0cm},clip,width=\linewidth]{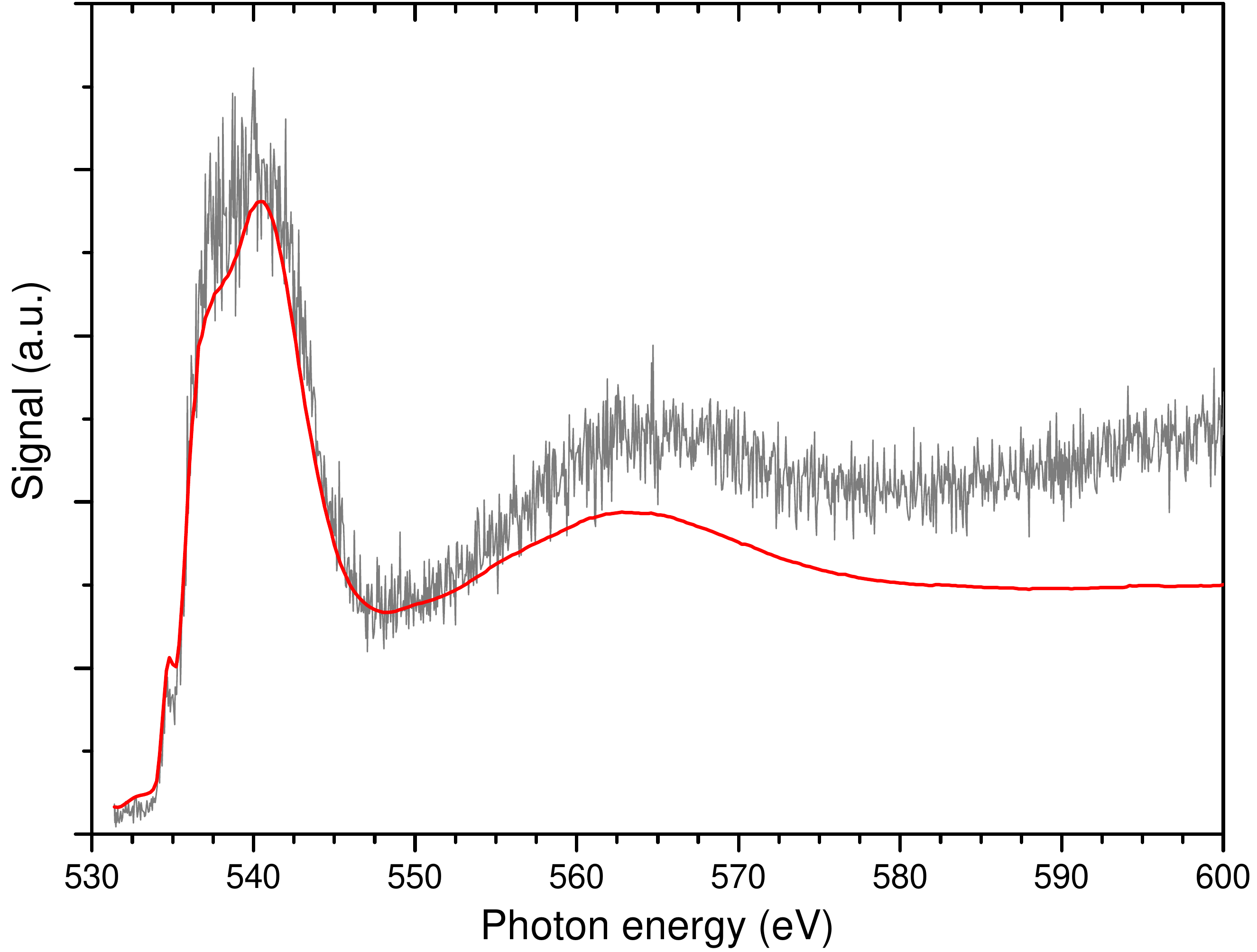}
    \caption{Photodesorption spectrum of (H$_2$O)$_3$H$^+$ at 90 K from c-ASW. Indicated in red is the TEY for comparison, normalized so as to best make the different features match.}
    \label{cluster_spec}
\end{figure} 

The photodesorption spectra of all protonated clusters are similar. An example is shown in figure \ref{cluster_spec} for (H$_2$O)$_3$H$^+$. The spectrum follows the TEY, except for a deviation at high energies where a rising slope seems to be added to the TEY part. Considering the similarities between protonated cluster desorption in ESD and X-ray core excitation, we could argue that XESD dominates the desorption of clusters. However, the case is different from the desorption of neutrals, for which we claimed that XESD dominates. As mentioned previously, desorption of neutrals by electrons has a threshold of around 6.5 eV, corresponding to "simple" valence excitations or ionizations and easily attained by secondary electrons. On the other hand, clusters desorption has an energy threshold by ESD of as high as 70 eV \rev{(n $\geq$ 2)}, corresponding to highly excited "complex" states\cite{herring-captain2005}. These cannot be reached by secondary electrons. The initial Auger electron can excite such states but not with a high probability. On the other hand, decay of the initially excited molecule naturally leads to this kind of highly excited state. Therefore in this case, there is no good argument favouring XESD, and \rev{direct desorption} should play an important role. This is also corroborated by the deviation from the TEY at high energy. It could be suggested that this deviation simply corresponds to a slowly increasing desorption probability following the increasing energy of the photoelectron from core ionisation. However, this is not coherent with the measured ESD threshold of cluster desorption at 70 eV\cite{herring-captain2005}. Considering a ionization potential (IP) at 537 eV\cite{baron1976}, the kinetic energy of the photoelectrons would reach 70 eV for a photon energy of 607 eV, which is above our photon energy range. This might instead be linked to multielectron excitations (i.e. excitation/ionisation of a valence electron alongside the core electron) embedded in the ionization continuum, that have a small excitation cross-section (and cannot be seen in the absorption spectrum) but yield final states with multiple holes that are (i) more likely to stay localized on a molecule and (ii) propitious to Coulomb explosion and cluster desorption. 

\subsection{Desorption of H$^+$}

\begin{figure}
    \includegraphics[trim={0cm 0cm 0cm 0cm},clip,width=\linewidth]{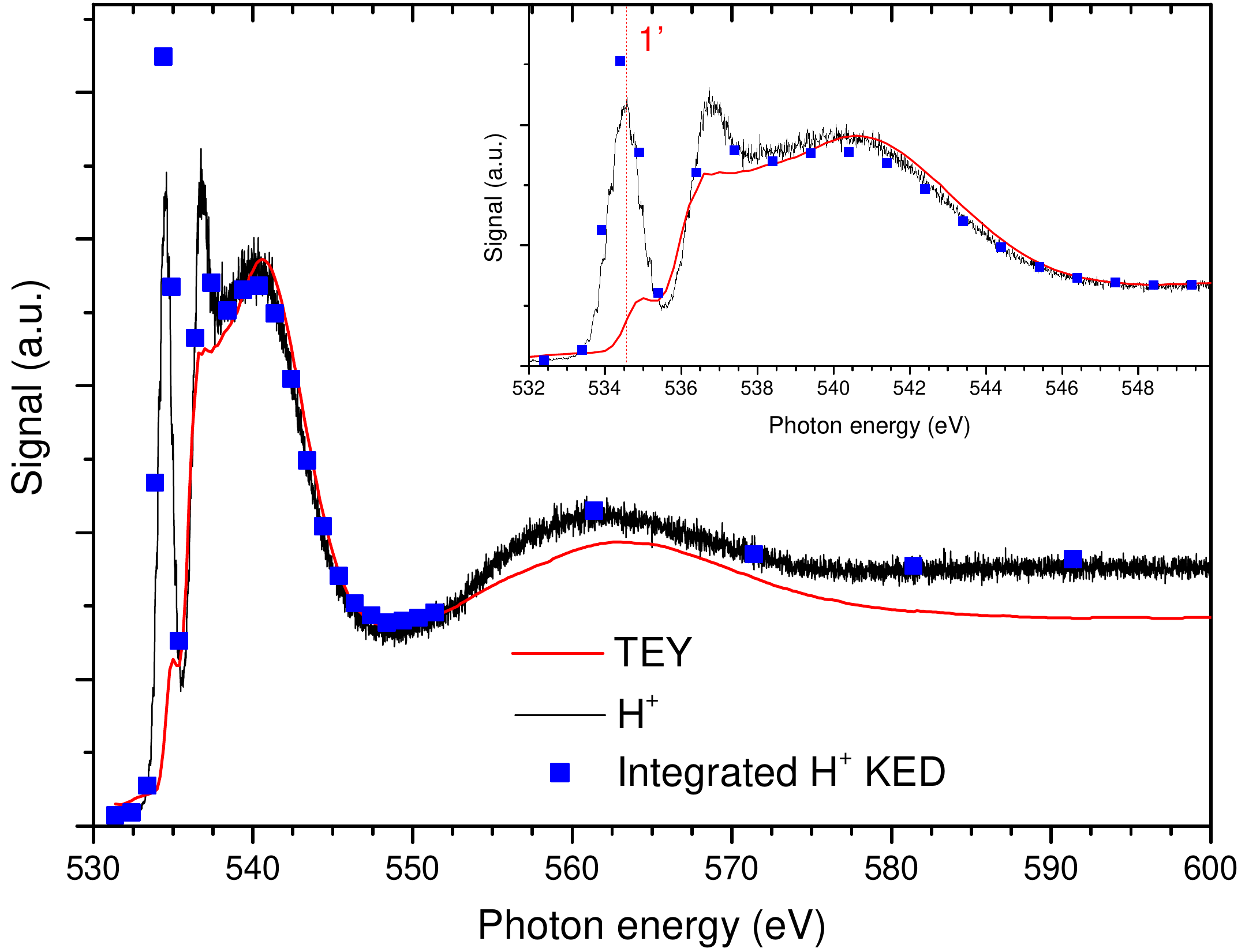}
    \caption{Photodesorption spectrum of H$^+$ for a 90 K c-ASW. The black trace is the spectrum at a kinetic energy of 7 eV, while the dots correspond to the signal integrated over the whole kinetic energy distribution (KED), showing that the spectrum at 7 eV is not distorted compared to the integrated signal. Also shown is the TEY (red trace) for comparison, normalized so as to best make the features match. Inset: zoom into the 532-550 eV region, showing the difference between peak 1 and peak 1'.}
    \label{H+}
\end{figure}

As mentioned in the introduction, the desorption of H$^+$ has been the most studied in water ice core excitation so far. Figure \ref{H+} shows the photodesorption spectrum of H$^+$ at the maximum of the kinetic energy distribution ($\sim$ 7 eV), along with points that correspond to the integrated kinetic energy distributions at various photon energies and a comparison with the TEY. We can see that the photodesorption spectrum shown here corresponds well to the result we obtain by integrating over the whole kinetic energy distribution. We have observed evidence that taking a spectrum at different kinetic energy settings can change the relative intensities of some features, which will not be discussed in this paper. 

Here we see that the desorption spectrum deviates from the TEY for several features. A prominent feature of the spectrum is the red-shifted and much more intense peak at 534.4 eV. We will label it peak 1' here. This feature has been studied in detail, \rev{including e.g. thickness and polarization-dependence studies}\cite{coulman1990}. As explained in the introduction the following conclusions were reached: (i) it originates from direct core dissociation of surface water molecules, those with a dangling H, and (ii) evidence suggests a fast dissociation mechanism, with significant nuclear motion already during the lifetime of the core hole. The surface origin of the feature is confirmed by surface-sentitive grazing incidence Auger electron spectroscopy\cite{nordlund2004}. The fast dissociation mechanism is further compounded by desorbing ion/Auger electron coincidence studies\cite{mase1998}, which reveals what final states are involved in desorption. 

\begin{figure}
    \includegraphics[trim={0cm 0cm 0cm 0cm},clip,width=\linewidth]{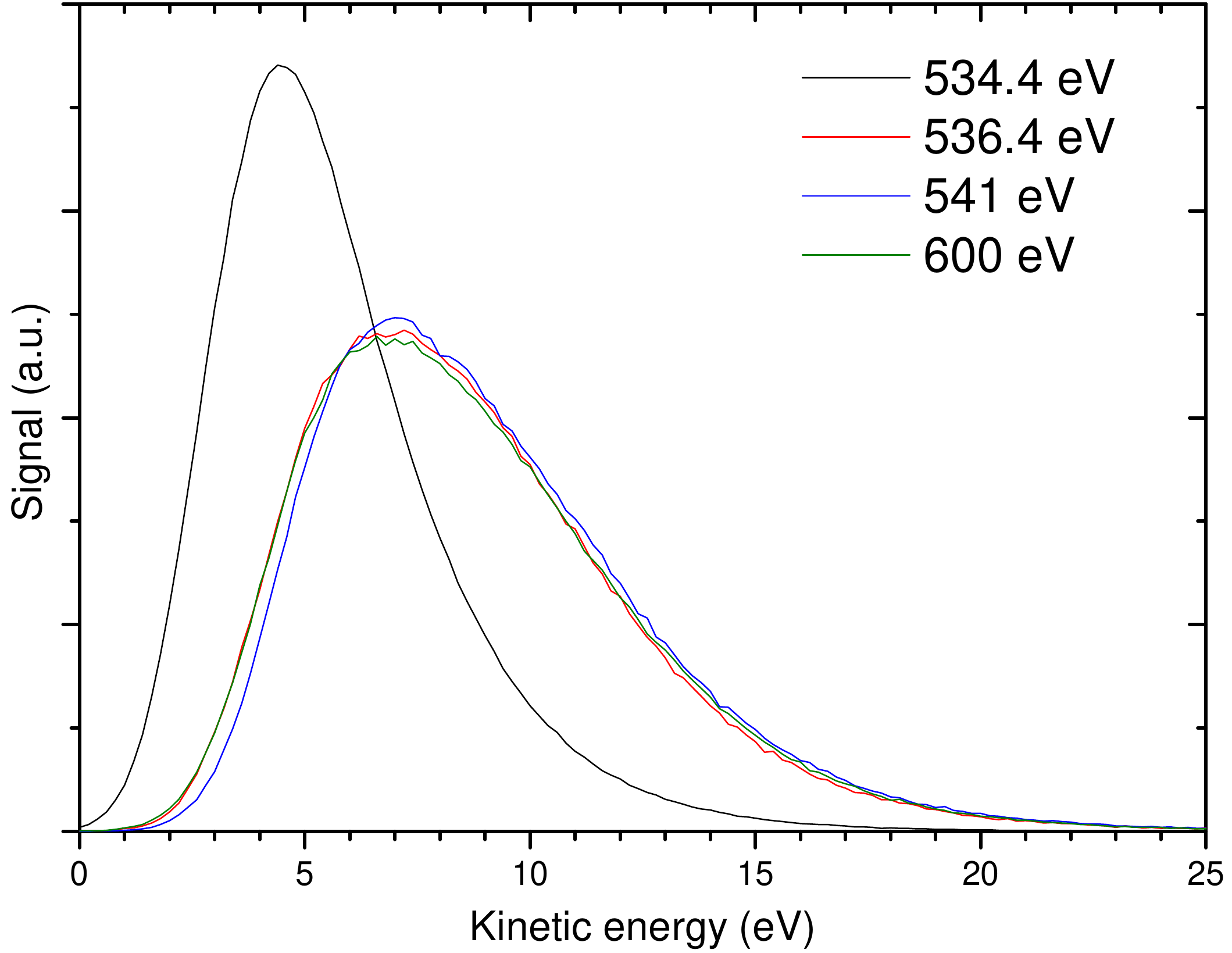}
    \caption{Kinetic energy distributions of H$^+$ ions at different photon energies, for c-ASW at 90 K.}
    \label{H+KE}
\end{figure}

We have measured the kinetic energy distribution of H$^+$ ions at various photon energies (fig. \ref{H+KE}). Some rough measurements had already been presented by Coulman et al.\cite{coulman1990}, who had remarked that ions created at the pre-edge feature were slower than others. Here we see that H$^+$ at the main edge excitation or above ionisation at 600 eV have the same kinetic energy distribution, while it is very different for ions at peak 1' (534.4 eV). The kinetic energy distribution at peak 1' is narrower and has a maximum at lower kinetic energy. This difference is well explained considering the fast dissociation mechanism. When exciting peak 1', the O-H bond is significantly elongated during the lifetime of the core hole, and the kinetic energy acquired by the proton is thus governed by the shape of the repulsive potential curve of the core-excited state along this coordinate. For other photon energies, various final states (2h1e and 2h states involving different valence orbitals) are at the origin of the proton fragments and the kinetic energy reflects the variety of these states (see ref \cite{sieger1997,ramaker1983b} for a detailed discussion of the kinetic energies for the different final states). 

Another feature is visible on the spectrum at 536.8 eV that resembles peak 1', in the sense that it is a sharp peak more intense than in the TEY, although the increase is not as dramatic as in the case of peak 1'. This feature has been suggested to correspond to a surface feature as well\cite{coulman1990}. It is however different from the previously discussed feature, as at this photon energy the kinetic energy distribution is similar to other photon energies (fig. \ref{H+KE}). This peak has a polarization dependence orthogonal to peak 1' according to the detailed polarization dependence studies of Coulman et al.\cite{coulman1990}. Coincidence studies around this energy show the expected results for regular Auger decay\cite{mase2003} (i.e. results similar to those obtained when probing "bulk" features), however the peak is not evident in their spectra, which might be due to the polarization they used.  Still, the conclusion is that there is no evidence of fast dissociation associated with this feature, which is however a \rev{direct desorption} feature. 

The H$^+$ spectrum also deviates from the TEY in the EXAFS region, from 550 eV and above. The EXAFS region, and the first EXAFS oscillation at around 560 eV, are characteristic of the O-O distance and arrangement inside the ice. Therefore it was previously suggested that the discrepancy between the H$^+$ and the TEY spectra could be attributed to a surface vs bulk behaviour as well\cite{coulman1990,parent2002}, with the surface molecules having higher average O-O distances. The difference derived from an EXAFS analysis by Parent et al.\cite{parent2002} was rather small, 2.76 \AA~for the bulk vs 2.77 \AA~for the surface. On the other hand, Coulman et al.\cite{coulman1990} had previously claimed that EXAFS analysis of the H$^+$ spectrum was not possible, due to the fact that in this region, contrary to peak 1', secondary electrons and bulk excitations still contribute significantly to the H$^+$ yield. The spectrum in this region is therefore more complicated to interpret. Aside from the peak shift, there is also an increase of the non-resonant continuum relative to the rest. It is worth noting that this increase is slightly different from the rise observed in the clusters spectra, but this does not preclude the possibility of a similar explanation: the role played by multi-electron excitations. Such states would indeed be expected to have higher proton fragment yields than double-ionized states, as they are more likely to stay localized on the molecule and lead to a Coulomb explosion. Multi-electron excitations have been shown to play an even more important role in desorption of ions after core ionization of adsorbed CO, for example\cite{feulner2000}.

\rev{Direct desorption} and surface molecules therefore clearly play an important role in H$^+$ desorption. However, outside of peak 1' it is difficult to separate the possible contributions of \rev{direct desorption} and XESD. It is even less likely that H$^+$ desorption is a surface-only probe in the whole region of the spectrum. 

\subsection{Desorption of H$^-$}

The desorption of anions in X-ray core excitation of condensed molecules has been observed before\cite{dujardin1989,andrade2010}, showing such highly ionizing radiation can still yield negative ions. Anions in the present case can come from two distinct mechanisms, once again related to the distinction between XESD and \rev{direct desorption}. Electrons can mainly create anions by \rev{dissociative electron attachment (DEA) or by ion pair dissociation\cite{sanche1990,bass2003}.} Low energy electron stimulated desorption of anions from water ice, especially through the DEA process, has been studied extensively\cite{tronc1996,simpson1997}. The DEA resonance of water around 7-9 eV corresponds well to the typical energy of secondary electrons, while the Auger electron and higher energy secondaries can cause ion pair dissociation from the neutral molecule. The other possibility for anion formation is direct dissociation of the core excited molecule in an anion fragment and highly charged cation fragments. Such dissociation pathways, although less likely than cation or cation-neutral dissociation, have been observed in the gas phase\cite{stolte2003,piancastelli2005,strahlman2016}. Their cross-section is typically 10$^{-3}$-10$^{-4}$ lower than dissociation pathways involving only cations and neutrals\cite{stolte2003}. 

\begin{figure}
    \includegraphics[trim={0cm 0cm 0cm 0cm},clip,width=\linewidth]{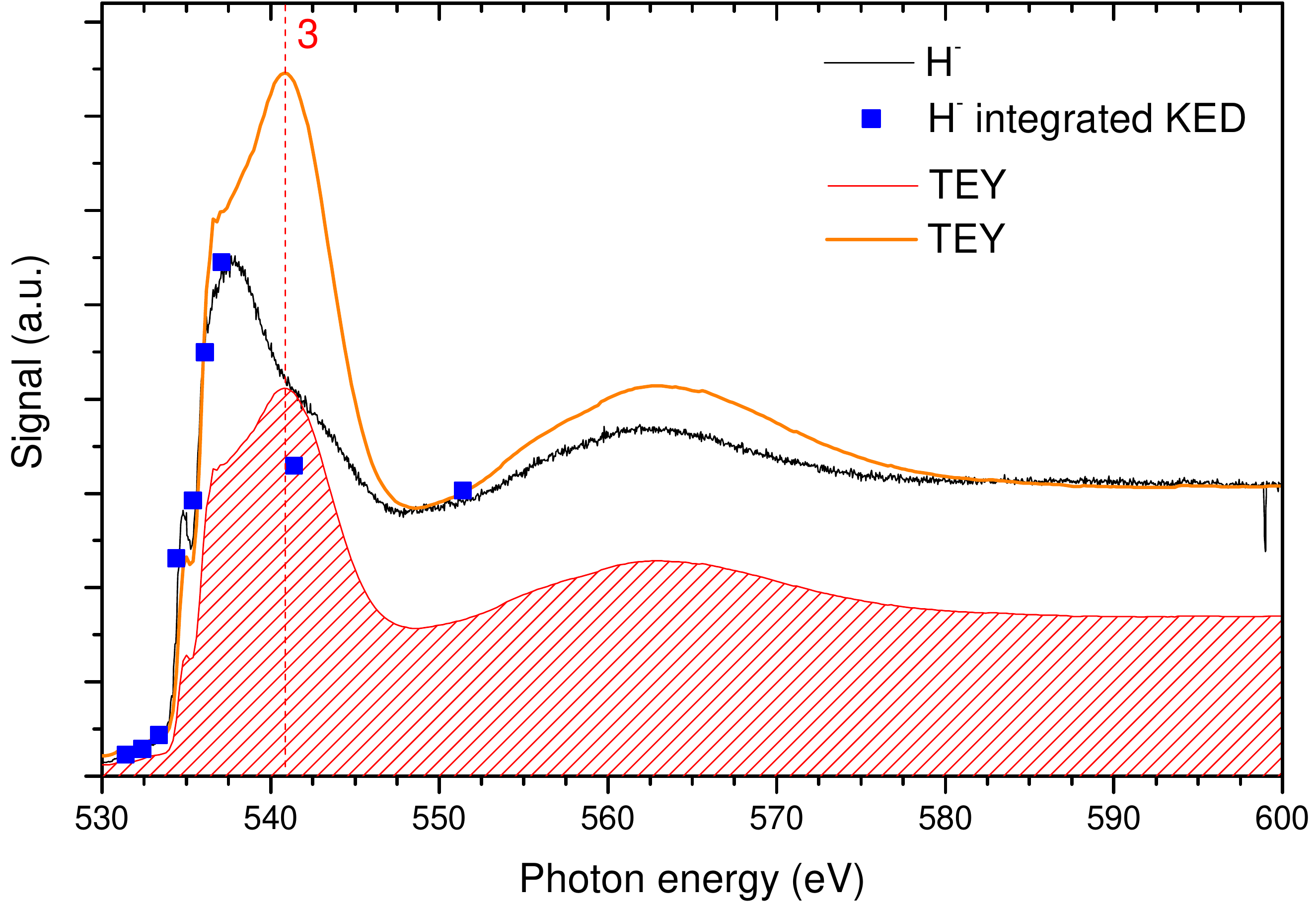}
    \caption{Photodesorption spectrum of H$^-$ at 90 K from c-ASW, at the maximum of the kinetic energy (black trace). The blue dots are the integrated H$^-$ KED. The TEY is indicated as an orange trace (normalized at 600 eV) and as a hatched red area under the H$^-$ desorption spectrum, representing the maximum contribution of XESD to H$^-$ desorption (see the text).}
    \label{H-}
\end{figure}

Fig \ref{H-} shows the photodesorption spectrum of H$^{-}$ at 90 K at the maximum of the kinetic energy distribution, along with a comparison with the TEY. The integrated KEDs for H$^-$ show a good match with the displayed spectrum, confirming its shape. The photodesorption spectrum differs markedly from the TEY around 541 eV, where the post-edge feature (peak 3) is strongly suppressed, and in the ionization continuum. The maximum contribution of XESD to the desorption of H$^{-}$ can be estimated from these differences. Indeed, the contribution of the XESD in the spectrum has to follow the spectral shape of the TEY. This maximum contribution corresponds to the hatched TEY-shaped area in red under the H$^-$ desorption spectrum. This hatched area is approximately 60\% of the total area. Conversely, \rev{direct desorption} processes in this spectrum therefore account for at least 40\% of the H$^{-}$ desorption. This quantification is only indicative as it depends on the energy range considered, but what we can conclude is that there has to be a \rev{direct desorption} process with an efficiency at least comparable to XESD. Such a result is surprising considering the orders of magnitude difference of cross-sections for the dissociation pathways involving or not an anion, mentioned above. We find an estimated one order of magnitude difference between desorption of H$^+$ and H$^-$ (assuming similar detection efficiencies for both ions). There is therefore a gap of at least two orders of magnitude with the gas phase results. One possibility is that the difference occurs at the desorption step, but it is difficult to see why H$^-$ would have a much higher desorption probability than H$^+$ after formation. \revd{H$^+$ can react with surrounding molecules, which is in competition with H$^+$ desorption, but so does H$^-$, which reacts with H$_2$O to form H$_2$ with a high cross-section\cite{orlando1999,kimmel1996}}. The more likely explanation is a condensed phase effect. All the possible rapid charge transfers to neighbouring molecules that can occur in the condensed phase seem to enhance significantly the probability of anion formation during core-excited/ionized water dissociation. 

On the H$^-$ desorption spectrum (fig. \ref{H-}), intensity is missing on peak 3, which as mentioned previously is attributed to four-coordinated molecules and a delocalized excitation. A delocalized excitation means the final state after Auger decay will a be a 2h state, as the electron density is quickly delocalized over several molecules. We could expect a 2h state to be less likely to yield anions than a 2h1e state, but then the H$^-$ signal in the ionization continuum would also be suppressed relative to the pre and main edge (peaks 1 and 2). This missing intensity may then rather be linked to the fact that mostly four-coordinated molecules are excited here, which for a reason yet to be found is less likely to yield anions. A similar explanation can also hold for the lessened EXAFS oscillation, i.e. a resonance also linked with the coordination shell of the molecules. 

\subsection{Desorption of oxygen-bearing ions}

The desorption of fragment cations other than H$^+$ was previously observed\cite{coulman1990} during core excitation with a much lower yield than for H$^+$ but no detailed studies were performed, and in particular no spectra were shown. The observation of low yields of oxygen-bearing fragments (O$^+$, OH$^+$, H$_2$O$^+$) desorbing from water ice is a common feature of not only soft X-ray irradiation\cite{coulman1990}, but also XUV irradiation\cite{rosenberg1981} and medium energy (100-200 eV) electrons\cite{herring-captain2005}. Studies of anion desorption by dissociative electron attachment (DEA) of low-energy ($<$20 eV) electrons on water ice also reported very low O$^-$ and OH$^-$ desorption signals\cite{pan2005}. Our estimated desorption yields for oxygen-bearing fragments are, similarly, much lower than H$^+$(table \ref{table_yields}). Several reasons can be invoked to explain these observations.

\begin{figure}
    \includegraphics[trim={0cm 0cm 0cm 0cm},clip,width=\linewidth]{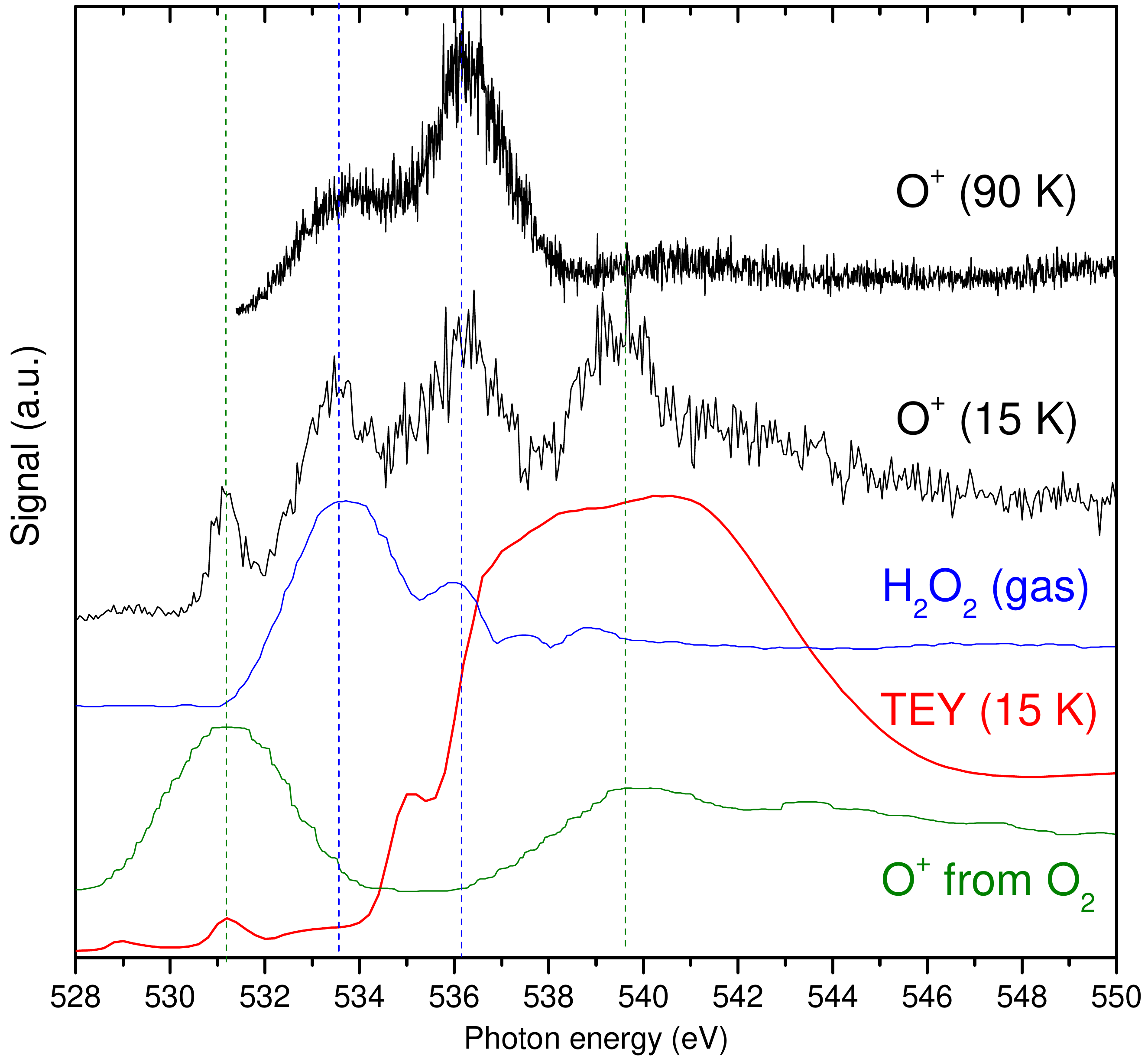}
    \caption{Photodesorption spectra of O$^+$ at 90 and 15 K from c-ASW. The TEY is indicated (red trace) for comparison, along with the gas-phase absorption spectrum of H$_2$O$_2$ (blue trace), adapted from Ruhl et al.\cite{ruhl1991} and shifted to match the first H$_2$O$_2$ peak, and the O$^+$ desorption spectrum for O$_2$ ice (green trace) from Rosenberg et al.\cite{rosenberg1985}}
    \label{O+}
\end{figure}

As the oxygen cation and anion fragments are readily observed in core dissociation of gas-phase water\cite{piancastelli1999,piancastelli2005,stolte2003}, we cannot ascribe the observed lower desorption of these fragments to a dissociation branching ratio that would favor neutral oxygen radicals instead of ions so heavily, especially in cases where multiple holes are involved. One important difference is the kinetic energy of the different fragments. In dissociation of an isolated water molecule into two fragments H and OH (regardless of their charge), momentum conservation implies that the H fragment takes away $\sim$94\% of the kinetic energy, and little is left for the oxygen-bearing fragment. In the case of dissociation into three fragments, the symmetry of the molecule also plays a role: it has been observed in Coulomb explosion of molecules that the central atom has much less kinetic energy because the recoil energy of the peripheric fragments compensate\cite{carlson1983}. Such would be the case for dissociation of H$_2$O in H + O + H. In a condensed medium there are also other channels for energy loss, as well as environment effects on the symmetry of the molecules, so that the theoretical kinetic energy partition in the isolated molecule is only an upper limit. The lower kinetic energy of the oxygen fragments would reduce their probability to overcome the desorption barrier. 

Other explanations have been suggested to explain this low desorption yield: one explanation is based on the assumption that the surface orientation of water molecules is always unfavourable to desorption of oxygen fragments\cite{rosenberg1981}. Another stems from the fact that there is a competition between desorption of an ion and decay of the ionic state through reneutralization\cite{ramaker1983b}. The H-bond network of ice facilitates rapid intermolecular charge transfer processes. This is in fact also linked with the previous argument on kinetic energies: fragments with lower kinetic energies will spend a longer time at or near the surface before escaping, which increases the likelihood of their reneutralization. Another possibility that would prevent desorption is if the cation is trapped. Once again, the trapping probability should be higher for slower ions, considering the time it would take for the structural rearrangement of surrounding dipoles that would stabilize the ion in the ice. Such a structural rearrangement certainly could not occur in the case of fast desorption of H$^+$, for example, which takes place on a timescale of a few tens of fs. 

\begin{figure}
    \includegraphics[trim={0cm 0cm 0cm 0cm},clip,width=\linewidth]{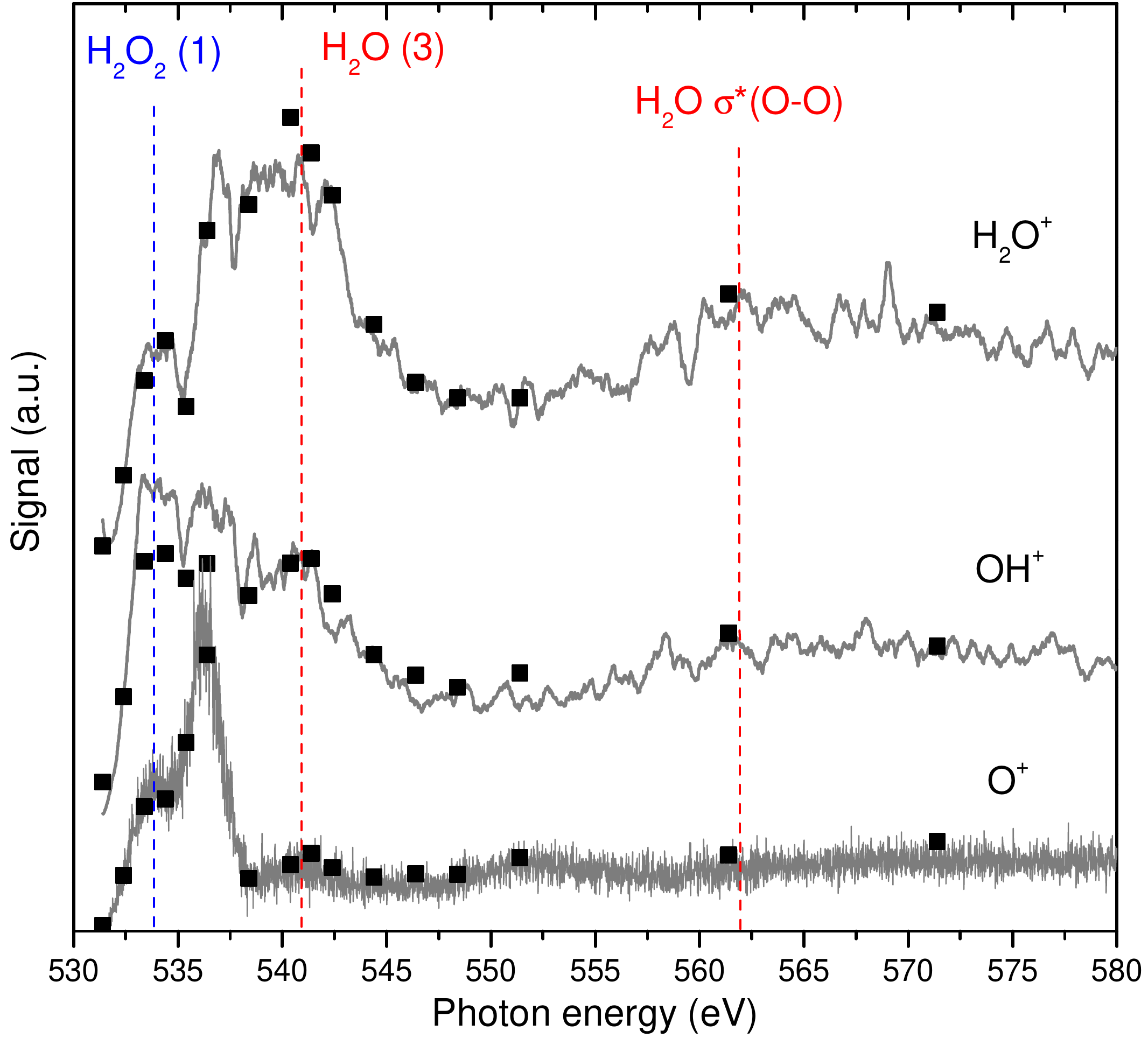}
    \caption{Photodesorption spectra of O$^+$, OH$^+$ and H$_2$O$^+$ at 90 K from c-ASW. The spectra were taken at the maxima of the kinetic energy distributions of the ions. The spectra of OH$^+$ and H$_2$O$^+$ have been smoothed. Also indicated are the integrated kinetic energy distributions at various photon energies (black dots) to show that these spectra are indeed representative of the non-differentiated spectra. The attribution of the main features are given, indicating the molecule and peak attribution, using the same color code as in fig. \ref{O+}. The curves have been arbitrarily scaled to have roughly the same size.}
    \label{Ox+}
\end{figure}

\begin{figure}
    \includegraphics[trim={0cm 0cm 0cm 0cm},clip,width=\linewidth]{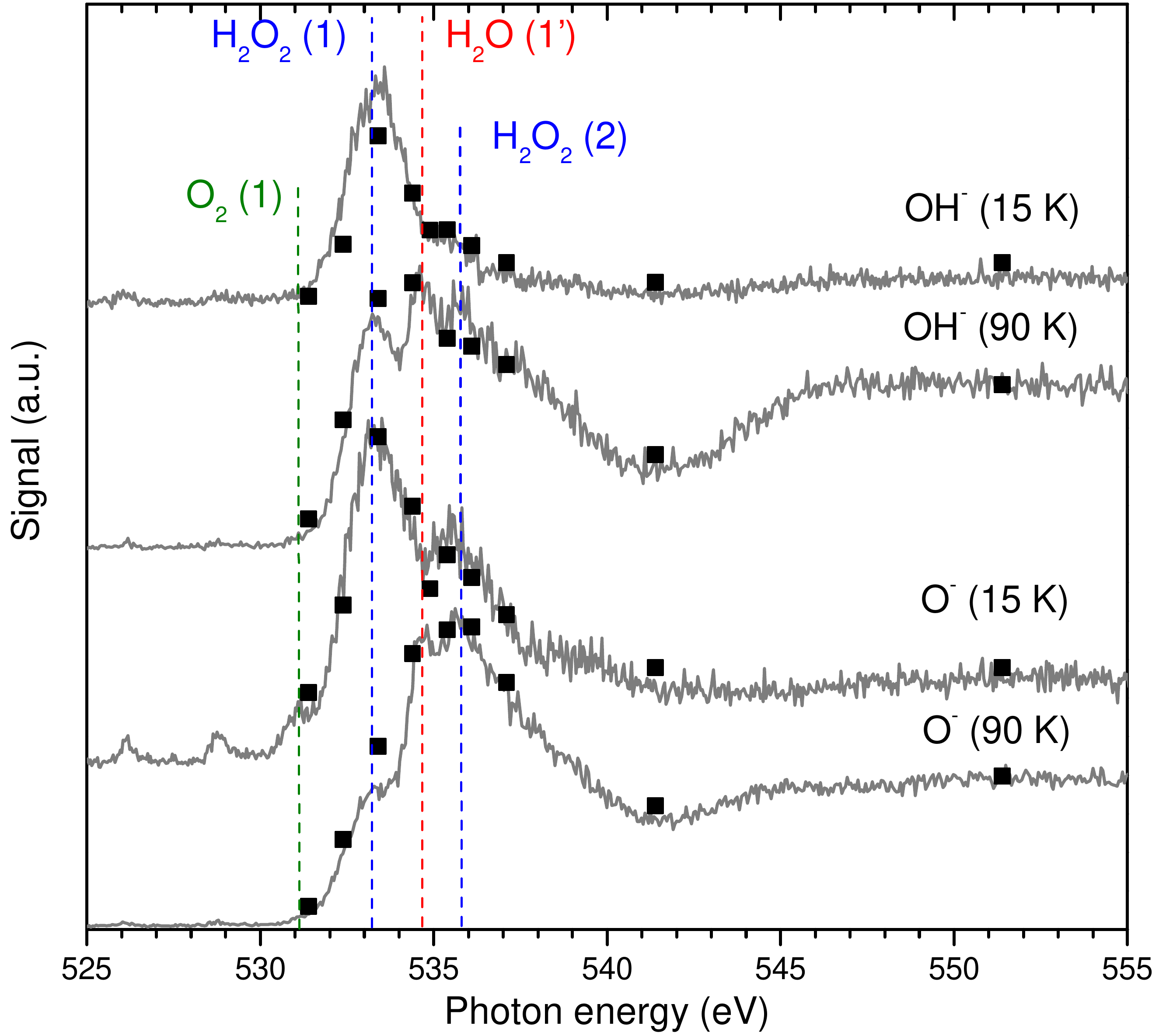}
    \caption{Photodesorption spectra of O$^-$ and OH$^-$ at 90 K or 15 K from c-ASW. The spectra were taken at the maxima of the kinetic energy distributions of the ions. Also indicated are the integrated kinetic energy distributions at various photon energies (black dots) to show that the these spectra are indeed representative of the non-differentiated spectra. The attribution of some peaks are given, with the molecule and the number of the peak, using the same color code as in fig. \ref{O+}. The curves have been arbitrarily scaled to have roughly the same size.}
    \label{Ox-}
\end{figure}

The desorption spectra of oxygen-bearing fragments provide useful information to discuss their low desorption efficiency. The desorption spectrum of O$^+$ in fig. \ref{O+} bears no resemblance to the TEY. Instead the spectrum at 90 K is dominated by two features which can be recognized as two peaks of the spectrum of H$_2$O$_2$. The first one is in fact visible in the TEY as mentioned in section B. The comparison to the gas phase H$_2$O$_2$ spectrum (O$^+$ yield\cite{ruhl1991}, shifted to match the first peak) shown in the figure allows to attribute the second feature, which is not visible in the TEY where it would overlap with and be hidden by the water peaks. Looking at the desorption spectrum of O$^+$ at 15 K (fig. \ref{O+}), two other features are observed which can be attributed to O$_2$ (the desorption of O$^+$ from O$_2$ ice from Rosenberg et al.\cite{rosenberg1983} is shown as well for comparison). Once again the first O$_2$ feature is visible in the TEY, although weak, and we saw that it was more abundant in the ice at 15 K than at 90 K. This spectral information indicates that the desorption of O$^+$ is dominated by direct excitation and dissociation (therefore a \rev{direct desorption} process) of products of the photochemistry, which are present only at the few percent level in the ice. We cannot exclude that the concentration of photoproducts is higher at the surface than in the bulk, because diffusion of radicals, and the subsequent chemistry, can be different at the surface. Still, considering the relative abundances, the desorption of O$^+$ through excitation of these photoproducts must be much more efficient than through XESD or direct excitation of H$_2$O. This strong contribution of H$_2$O$_2$ is also seen in the desorption spectra of the other oxygen-bearing fragments, OH$^+$, H$_2$O$^+$ (fig. \ref{Ox+}), O$^-$ and OH$^-$ (fig. \ref{Ox-}).

These observations are coherent with the idea that kinetic energy partition during dissociation is important. Indeed, dissociation of H$_2$O$_2$ or O$_2$ instead of H$_2$O implies breaking (at least) an O-O bond, yielding fragments of similar masses and thus releasing more energy into oxygen-bearing fragments. It would also at first weaken the explanation of a competition between desorption and decay of an ionic state, as there is no reason that an O$^+$ or OH$^+$ fragment from H$_2$O$_2$ would have a different decay probability than if they resulted from H$_2$O dissociation. However this is not completely ruled out because, as pointed out above the explanations can be complementary rather than contradictory: a higher kinetic energy for the fragment implies less time spent in the ice and thus a lower probability of decay. Another consideration is that we do not know the structural arrangement of photoproducts in the ice: if the connection of these species to the H-bonding network is poor, the lifetime of the ionic states could be longer and also favor desorption. Similarly if there are species with dangling OH or O pointing outwards at the surface desorption would be favored. The various factors explaining low desorption of O-bearing fragments from ice (kinetic energy partition from dissociation, reneutralization probability, structural arrangement...) are inter-connected, thus it is difficult to rule out one of them, but the present experimental observations are consistent with the general picture they form.  

Comparisons of the spectra of the different fragments also show some interesting differences. In fig. \ref{Ox+}, the spectra of O$^+$, OH$^+$ and H$_2$O$^+$ for a 90 K ice are plotted. The integrated kinetic energy distributions at different photon energies are plotted as well (black dots) to show that the displayed spectra are indeed representative of the non-differentiated spectra. We can see that the relative contributions to the desorption of H$_2$O$_2$ and H$_2$O (either XESD or \rev{direct desorption}), as indicated by the relative intensities of the first H$_2$O$_2$ peak and the H$_2$O peak 3 and $\sigma$*(O-O), are different for the three ions. The spectrum of O$^+$ is devoid of any clear H$_2$O features. On the other hand, the H$_2$O peak 3 at 541 eV and the $\sigma$*(O-O) at 562 eV are visible in the spectrum of OH$^+$, although strong contributions of H$_2$O$_2$ are still present. In the H$_2$O$^+$ spectrum, the H$_2$O contributions become stronger than the H$_2$O$_2$ contributions.

The difference between O$^+$ and OH$^+$ would be in line with the symmetry considerations mentioned above: in dissociation of H$_2$O in H + H + O$^+$ (regardless of the charge of the H fragments), not only is the kinetic energy taken away by the lighter H fragments, the central place of the O atom would also yield it even less kinetic energy than in the case of a dissociation in two fragments. This could explain the absence of any water-related feature in the O$^+$ spectrum. The case of H$_2$O$^+$ is different: H$_2$O$^+$ is the result of simple ionization of water, with no dissociation is involved. Simple ionization is unlikely to lead to H$_2$O$^+$ desorption, considering there should be a rather high desorption barrier due to the stability of the ion in a water matrix (and also the competition with formation of H$_3$O$^+$, presumably). A mechanism should therefore be invoked to explain the contribution of H$_2$O to the yield. One possibility is that the energy for desorption comes from Coulomb repulsion between two neighbour ionized molecules. Such a situation can happen when a charge transfer or proton transfer occurs from a 2h final state, or when the core ionized state undergoes ICD. If we consider such a mechanism to be at the origin of H$_2$O$^+$ desorption after H$_2$O excitation, it is more likely that \rev{direct desorption} will dominate over XESD, since double-ionization processes are not energetically possible for secondary electrons and not very likely for Auger electron scattering. \rev{This mechanism is similar to the one of cluster desorption. In fact it is probably a minor outcome channel of the process that can also lead to cluster desorption. When H$_2$O$^+$ forms it is expected that proton transfer will form H$_3$O$^+$ instead - rapidly: the time scale for proton transfer typically competes with ICD. H$_2$O$^+$ desorption can occur when desorption outpaces proton transfer, a presumably minor case compared to H$_3$O$^+$ or cluster formation and desorption, which is reflected in the desorption yield being an order of magnitude lower (Table \ref{table_yields}).}

\begin{figure}
    \includegraphics[trim={0cm 0cm 0cm 0cm},clip,width=\linewidth]{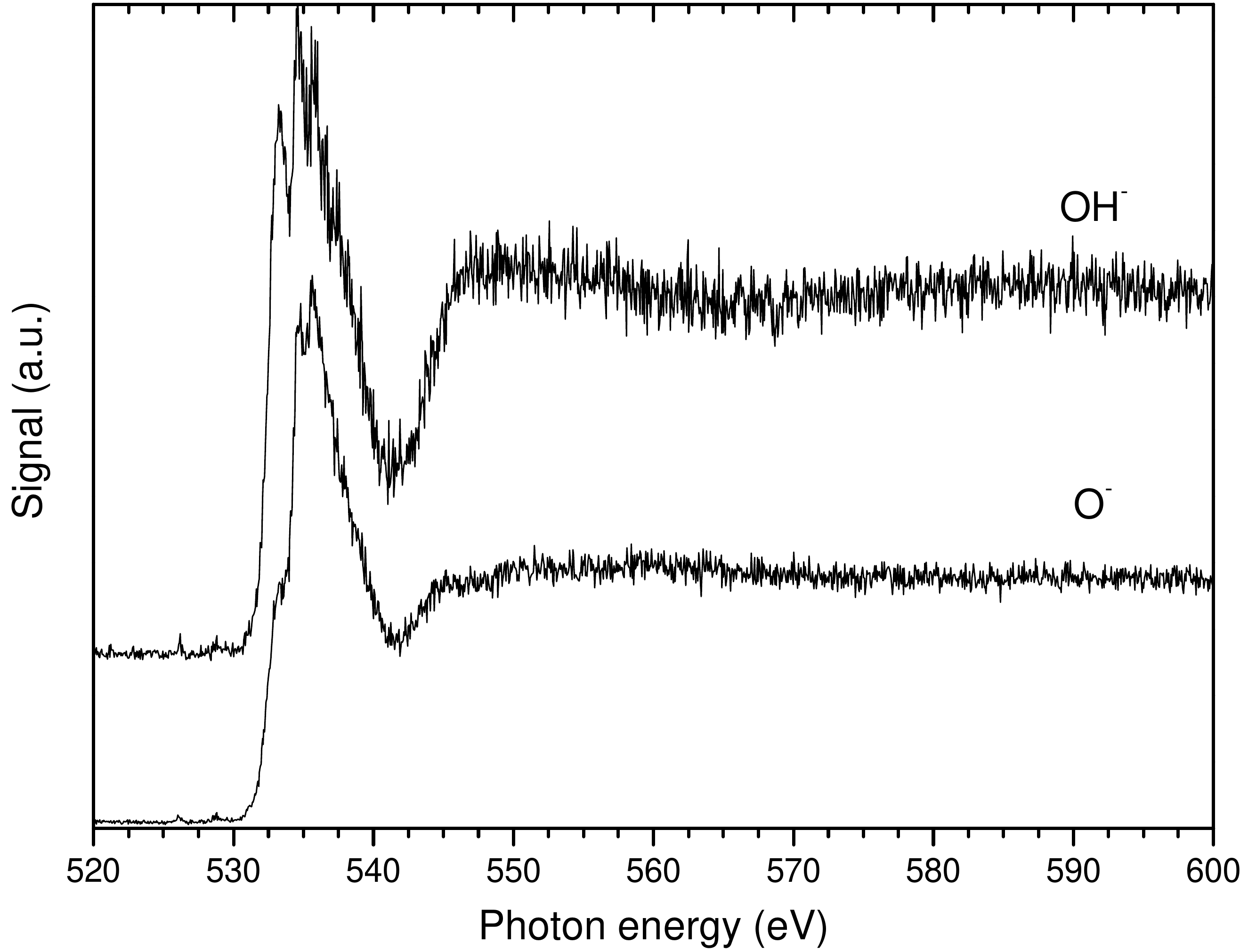}
    \caption{Photodesorption spectra of O$^-$ and OH$^-$ at 90 K from c-ASW, from 520 to 600 eV. The spectra were taken at the maxima of the kinetic energy distributions of the ions. The curves have been arbitrarily scaled to have roughly the same size.}
    \label{Ox-_full}
\end{figure}

The desorption spectra of the anions O$^-$ and OH$^-$ are shown in the edge region in figure \ref{Ox-} (at 15 and 90 K) and extended up to 600 eV in figure \ref{Ox-_full} (at 90 K only). The H$_2$O$_2$ features are prominent in the edge region while the bulk water features such as peak 3 at 541 eV and the $\sigma$*(O-O) resonance at 560 eV are completely absent, similar to O$^+$ (fig. \ref{O+}), although the spectra are still different in the continuum region. However, surprisingly, a peak corresponding to peak 1' of water at  534.4 eV is observed, especially at 90 K (at 15 K it could be hidden by the H$_2$O$_2$ peaks). The presence of this peak in the absence of any other water feature is surprising and suggests that O$^-$ and OH$^-$ anions may be produced in the fast desorption process discussed for H$^+$. This is especially surprising in the case of OH$^-$, since formation of OH$^-$ from water dissociation after core excitation necessarily implies either charge transfer to neighbouring molecules or radiative, rather than Auger decay (1\% of the events). More investigations would be required to see if this apparent feature can truly be attributed to H$_2$O. Another feature that is distinctive of these two ions is the ionization continuum (fig. \ref{Ox-_full}, above $\sim$540 eV), for which we see a clear sigmoid shape with no EXAFS feature or shape resonance. A possible attribution of this feature is the ionization continuum of H$_2$O$_2$. However, the fact that it is not seen for O$^+$ raises doubt about this interpretation. The lack of resemblance with the TEY of these continuum features rules out an XESD contribution. A \rev{direct desorption} contribution from H$_2$O excitation is also unlikely, since no dissociation pathway starting from excited H$_2$O$^{2+}$ (after core ionization and Auger decay) could yield O$^-$ or OH$^-$. Such a \rev{direct desorption} process would only be possible, again, if transfer of a positive charge to a neighbouring molecule occurs, or after radiative decay.

\section{Conclusion}

The objective of this study was to explore all the species that are released in the gas phase when core-exciting amorphous water ice, to attempt to distinguish different desorption mechanisms when possible, and to see what information can the desorption of some specific species tell us about dissociation, chemistry and relaxation following core-excitation. We have observed neutral species, cations and anions (summarized in table \ref{table_yields}), and the escaping electrons as well. We quantified the desorption of neutrals and the escaping electrons, and tentatively quantified the desorption of ions. 

We conclude that XESD - desorption mediated by the Auger and secondary electrons - is dominant in neutral desorption. For ions, there is evidence that this is not the case at all photon energies and \rev{direct desorption} processes - desorption mediated directly by the core-excited/ionized molecule - contribute as well. XESD could play a role in the desorption of H$^+$ or protonated clusters, but it is much less clear than for neutrals. \rev{Direct desorption} is at least as efficient as XESD in the case of H$^-$ desorption, so that there must exist dissociation pathways of core-excited and even core-ionized H$_2$O that lead to anions, in addition to the processes induced by secondary electrons. These dissociation pathways are identified in the gas phase but are not very efficient, so the very clear signals we obtain for ion desorption are still somewhat surprising and suggest a strong condensed phase effect (due to the possibility of charge transfers) favouring anion formation. We have also shown that \rev{direct desorption} dominates completely O$^+$ desorption and always plays a major role in the desorption of other oxygen fragment ions (OH$^+$, H$_2$O$^+$, O$^-$, OH$^-$). An important conclusion for these oxygen fragments is that \rev{direct desorption} following core-excitation of water itself is not efficient: instead \rev{direct desorption} following core-excitation of radiolysis products and in particular H$_2$O$_2$ plays a very important role. 

Now after this broad exploration, efforts should focus on the details of the desorption of specific species such as H$^-$ or O$^+$, similarly to the work that has been done on H$^+$, to further understand the information they can yield on core-excitation of water ice, since clearly interesting elements remain to be uncovered. 

\section*{Acknowledgments}  

\textit{This work was done in collaboration and with financial support by the European Organization for Nuclear Research (CERN) under the collaboration agreement KE3324/TE. We acknowledge SOLEIL for provision of synchrotron radiation facilities under the projects 20161406 and 20181140 and we thank Nicolas Jaouen and the SEXTANTS team for their help on the beamline. This work was supported by the Programme National "Physique et Chimie du Milieu Interstellaire" (PCMI) of CNRS/INSU with INC/INP co-funded by CEA and CNES. Financial support from the LabEx MiChem, part of the French state funds managed by the ANR within the investissements d'avenir program under reference ANR-11-10EX-0004-02, and by the Ile-de-France region DIM ACAV program, is gratefully acknowledged.} 

% Create the reference section using BibTeX:
\bibliographystyle{naturemag}
\bibliography{H2O_X_chph}

\begin{thebibliography}{10}
\expandafter\ifx\csname url\endcsname\relax
  \def\url#1{\texttt{#1}}\fi
\expandafter\ifx\csname urlprefix\endcsname\relax\def\urlprefix{URL }\fi
\providecommand{\bibinfo}[2]{#2}
\providecommand{\eprint}[2][]{\url{#2}}

\bibitem{vandishoeck2013}
\bibinfo{author}{{van Dishoeck}, E.~F.}, \bibinfo{author}{Herbst, E.} \&
  \bibinfo{author}{Neufeld, D.~A.}
\newblock \bibinfo{title}{Interstellar {{Water Chemistry}}: {{From Laboratory}}
  to {{Observations}}}.
\newblock \emph{\bibinfo{journal}{Chemical Reviews}}
  \textbf{\bibinfo{volume}{113}}, \bibinfo{pages}{9043--9085}
  (\bibinfo{year}{2013}).

\bibitem{vandishoeck2014c}
\bibinfo{author}{{van Dishoeck}, E.~F.}, \bibinfo{author}{Bergin, E.~A.},
  \bibinfo{author}{Lis, D.~C.} \& \bibinfo{author}{Lunine, J.~I.}
\newblock \bibinfo{title}{Water: {{From Clouds}} to {{Planets}}}.
\newblock In \emph{\bibinfo{booktitle}{Protostars and {{Planets VI}}}}
  (\bibinfo{publisher}{{University of Arizona Press}}, \bibinfo{year}{2014}).

\bibitem{grannas2007}
\bibinfo{author}{Grannas, A.~M.} \emph{et~al.}
\newblock \bibinfo{title}{An overview of snow photochemistry: Evidence,
  mechanisms and impacts}.
\newblock \emph{\bibinfo{journal}{Atmos. Chem. Phys.}} \bibinfo{pages}{45}
  (\bibinfo{year}{2007}).

\bibitem{dupuy2018}
\bibinfo{author}{Dupuy, R.} \emph{et~al.}
\newblock \bibinfo{title}{X-ray photodesorption from water ice in
  protoplanetary disks and {{X}}-ray-dominated regions}.
\newblock \emph{\bibinfo{journal}{Nature Astronomy}}
  \textbf{\bibinfo{volume}{2}}, \bibinfo{pages}{796--801}
  (\bibinfo{year}{2018}).

\bibitem{hogerheijde2011}
\bibinfo{author}{Hogerheijde, M.~R.} \emph{et~al.}
\newblock \bibinfo{title}{Detection of the water reservoir in a forming
  planetary system}.
\newblock \emph{\bibinfo{journal}{Science}} \textbf{\bibinfo{volume}{334}},
  \bibinfo{pages}{338--340} (\bibinfo{year}{2011}).

\bibitem{kamp2013}
\bibinfo{author}{Kamp, I.} \emph{et~al.}
\newblock \bibinfo{title}{Uncertainties in water chemistry in disks: {{An}}
  application to {{TW Hydrae}}}.
\newblock \emph{\bibinfo{journal}{Astronomy \& Astrophysics}}
  \textbf{\bibinfo{volume}{559}}, \bibinfo{pages}{A24} (\bibinfo{year}{2013}).

\bibitem{mitchell2013}
\bibinfo{author}{Mitchell, E.} \emph{et~al.}
\newblock \bibinfo{title}{Ultraviolet photodesorption as a driver of water
  migration on the lunar surface}.
\newblock \emph{\bibinfo{journal}{Planetary and Space Science}}
  \textbf{\bibinfo{volume}{89}}, \bibinfo{pages}{42--46}
  (\bibinfo{year}{2013}).

\bibitem{garrett2005}
\bibinfo{author}{Garrett, B.~C.} \emph{et~al.}
\newblock \bibinfo{title}{Role of {{Water}} in {{Electron}}-{{Initiated
  Processes}} and {{Radical Chemistry}}: {{Issues}} and {{Scientific
  Advances}}}.
\newblock \emph{\bibinfo{journal}{Chemical Reviews}}
  \textbf{\bibinfo{volume}{105}}, \bibinfo{pages}{355--390}
  (\bibinfo{year}{2005}).

\bibitem{alizadeh2012}
\bibinfo{author}{Alizadeh, E.} \& \bibinfo{author}{Sanche, L.}
\newblock \bibinfo{title}{Precursors of {{Solvated Electrons}} in
  {{Radiobiological Physics}} and {{Chemistry}}}.
\newblock \emph{\bibinfo{journal}{Chemical Reviews}}
  \textbf{\bibinfo{volume}{112}}, \bibinfo{pages}{5578--5602}
  (\bibinfo{year}{2012}).

\bibitem{nilsson2010}
\bibinfo{author}{Nilsson, A.} \emph{et~al.}
\newblock \bibinfo{title}{X-ray absorption spectroscopy and {{X}}-ray {{Raman}}
  scattering of water and ice; an experimental view}.
\newblock \emph{\bibinfo{journal}{Journal of Electron Spectroscopy and Related
  Phenomena}} \textbf{\bibinfo{volume}{177}}, \bibinfo{pages}{99--129}
  (\bibinfo{year}{2010}).

\bibitem{bartels-rausch2012}
\bibinfo{author}{{Bartels-Rausch}, T.} \emph{et~al.}
\newblock \bibinfo{title}{Ice structures, patterns, and processes: {{A}} view
  across the icefields}.
\newblock \emph{\bibinfo{journal}{Reviews of Modern Physics}}
  \textbf{\bibinfo{volume}{84}}, \bibinfo{pages}{885--944}
  (\bibinfo{year}{2012}).

\bibitem{palmer2018}
\bibinfo{author}{Palmer, J.~C.}, \bibinfo{author}{Poole, P.~H.},
  \bibinfo{author}{Sciortino, F.} \& \bibinfo{author}{Debenedetti, P.~G.}
\newblock \bibinfo{title}{Advances in {{Computational Studies}} of the
  {{Liquid}}\textendash{{Liquid Transition}} in {{Water}} and {{Water}}-{{Like
  Models}}}.
\newblock \emph{\bibinfo{journal}{Chemical Reviews}}
  \textbf{\bibinfo{volume}{118}}, \bibinfo{pages}{9129--9151}
  (\bibinfo{year}{2018}).

\bibitem{coulman1990}
\bibinfo{author}{Coulman, D.} \emph{et~al.}
\newblock \bibinfo{title}{Excitation, deexcitation, and fragmentation in the
  core region of condensed and adsorbed water}.
\newblock \emph{\bibinfo{journal}{The Journal of Chemical Physics}}
  \textbf{\bibinfo{volume}{93}}, \bibinfo{pages}{58--75}
  (\bibinfo{year}{1990}).

\bibitem{herring-captain2005}
\bibinfo{author}{{Herring-Captain}, J.} \emph{et~al.}
\newblock \bibinfo{title}{Low-energy (5 \textendash{} 250 {{eV}})
  electron-stimulated desorption of {{H}}+, {{H2}}+, and {{H}}+({{H2O}})n from
  low-temperature water ice surfaces}.
\newblock \emph{\bibinfo{journal}{Physical Review B}}
  \textbf{\bibinfo{volume}{72}} (\bibinfo{year}{2005}).

\bibitem{rosenberg1981}
\bibinfo{author}{Rosenberg, R.~A.} \emph{et~al.}
\newblock \bibinfo{title}{The photodissociative ionization of amorphous ice}.
\newblock \emph{\bibinfo{journal}{Chemical Physics Letters}}
  \textbf{\bibinfo{volume}{80}}, \bibinfo{pages}{488--494}
  (\bibinfo{year}{1981}).

\bibitem{rosenberg1983}
\bibinfo{author}{Rosenberg, R.~A.} \emph{et~al.}
\newblock \bibinfo{title}{K-shell excitation of {{D2O}} and {{H2O}} ice:
  {{Photoion}} and photoelectron yields}.
\newblock \emph{\bibinfo{journal}{Physical Review B}}
  \textbf{\bibinfo{volume}{28}}, \bibinfo{pages}{3026} (\bibinfo{year}{1983}).

\bibitem{parent2002}
\bibinfo{author}{Parent, P.}, \bibinfo{author}{Laffon, C.},
  \bibinfo{author}{Mangeney, C.}, \bibinfo{author}{Bournel, F.} \&
  \bibinfo{author}{Tronc, M.}
\newblock \bibinfo{title}{Structure of the water ice surface studied by x-ray
  absorption spectroscopy at the {{O K}}-edge}.
\newblock \emph{\bibinfo{journal}{The Journal of Chemical Physics}}
  \textbf{\bibinfo{volume}{117}}, \bibinfo{pages}{10842--10851}
  (\bibinfo{year}{2002}).

\bibitem{mase2003}
\bibinfo{author}{Mase, K.}, \bibinfo{author}{Nagasono, M.},
  \bibinfo{author}{Tanaka, S.-i.}, \bibinfo{author}{Sekitani, T.} \&
  \bibinfo{author}{Nagaoka, S.-i.}
\newblock \bibinfo{title}{Ion desorption from molecules condensed at low
  temperature: {{A}} study with electron-ion coincidence spectroscopy combined
  with synchrotron radiation ({{Review}})}.
\newblock \emph{\bibinfo{journal}{Low Temperature Physics}}
  \textbf{\bibinfo{volume}{29}}, \bibinfo{pages}{243--258}
  (\bibinfo{year}{2003}).

\bibitem{romberg2000}
\bibinfo{author}{Romberg, R.}, \bibinfo{author}{Frigo, S.},
  \bibinfo{author}{Ogurtsov, A.}, \bibinfo{author}{Feulner, P.} \&
  \bibinfo{author}{Menzel, D.}
\newblock \bibinfo{title}{Photon stimulated desorption of neutral hydrogen
  atoms from condensed water and ammonia by resonant {{O1s}} and {{N1s}}
  excitation: Search for the signature of ultrafast bond breaking}.
\newblock \emph{\bibinfo{journal}{Surface Science}}
  \textbf{\bibinfo{volume}{451}}, \bibinfo{pages}{116--123}
  (\bibinfo{year}{2000}).

\bibitem{weimar2011}
\bibinfo{author}{Weimar, R.}, \bibinfo{author}{Romberg, R.},
  \bibinfo{author}{Naydenov, B.}, \bibinfo{author}{Menzel, D.} \&
  \bibinfo{author}{Feulner, P.}
\newblock \bibinfo{title}{Dynamics of ultrafast dissociation of hydrogenic
  molecules by resonant antibonding core electron excitation: {{Conclusions}}
  from detuning dependence of {{H}}+- kinetic energy distributions from
  condensed films}.
\newblock \emph{\bibinfo{journal}{Chemical Physics Letters}}
  \textbf{\bibinfo{volume}{510}}, \bibinfo{pages}{78--81}
  (\bibinfo{year}{2011}).

\bibitem{desimone2013}
\bibinfo{author}{DeSimone, A.~J.}, \bibinfo{author}{Crowell, V.~D.},
  \bibinfo{author}{Sherrill, C.~D.} \& \bibinfo{author}{Orlando, T.~M.}
\newblock \bibinfo{title}{Mechanisms of {{H2O}} desorption from amorphous solid
  water by 157-nm irradiation: {{An}} experimental and theoretical study}.
\newblock \emph{\bibinfo{journal}{The Journal of chemical physics}}
  \textbf{\bibinfo{volume}{139}}, \bibinfo{pages}{164702}
  (\bibinfo{year}{2013}).

\bibitem{yabushita2013}
\bibinfo{author}{Yabushita, A.}, \bibinfo{author}{Hama, T.} \&
  \bibinfo{author}{Kawasaki, M.}
\newblock \bibinfo{title}{Photochemical reaction processes during
  vacuum-ultraviolet irradiation of water ice}.
\newblock \emph{\bibinfo{journal}{Journal of Photochemistry and Photobiology C:
  Photochemistry Reviews}} \textbf{\bibinfo{volume}{16}},
  \bibinfo{pages}{46--61} (\bibinfo{year}{2013}).

\bibitem{kimmel2001}
\bibinfo{author}{Kimmel, G.~A.}, \bibinfo{author}{Stevenson, K.~P.},
  \bibinfo{author}{Dohn{\'a}lek, Z.}, \bibinfo{author}{Smith, R.~S.} \&
  \bibinfo{author}{Kay, B.~D.}
\newblock \bibinfo{title}{Control of amorphous solid water morphology using
  molecular beams. {{I}}. {{Experimental}} results}.
\newblock \emph{\bibinfo{journal}{The Journal of Chemical Physics}}
  \textbf{\bibinfo{volume}{114}}, \bibinfo{pages}{5284--5294}
  (\bibinfo{year}{2001}).

\bibitem{hamers1998}
\bibinfo{author}{Hamers, E.}, \bibinfo{author}{{van Sark}, W.},
  \bibinfo{author}{Bezemer, J.}, \bibinfo{author}{Goedheer, W.} \&
  \bibinfo{author}{{van der Weg}, W.}
\newblock \bibinfo{title}{On the transmission function of an ion-energy and
  mass spectrometer}.
\newblock \emph{\bibinfo{journal}{International Journal of Mass Spectrometry
  and Ion Processes}} \textbf{\bibinfo{volume}{173}}, \bibinfo{pages}{91--98}
  (\bibinfo{year}{1998}).

\bibitem{rosenberg1985}
\bibinfo{author}{Rosenberg, R.~A.}, \bibinfo{author}{Love, P.~J.},
  \bibinfo{author}{LaRoe, P.~R.}, \bibinfo{author}{Rehn, V.} \&
  \bibinfo{author}{Parks, C.~C.}
\newblock \bibinfo{title}{K-shell photoexcitation of solid {{N2}}, {{CO}},
  {{NO}}, {{O2}}, and {{N2O}}}.
\newblock \emph{\bibinfo{journal}{Physical Review B}}
  \textbf{\bibinfo{volume}{31}}, \bibinfo{pages}{2634} (\bibinfo{year}{1985}).

\bibitem{timneanu2004}
\bibinfo{author}{T\i\^mneanu, N.}, \bibinfo{author}{Caleman, C.},
  \bibinfo{author}{Hajdu, J.} \& \bibinfo{author}{{van der Spoel}, D.}
\newblock \bibinfo{title}{Auger electron cascades in water and ice}.
\newblock \emph{\bibinfo{journal}{Chemical Physics}}
  \textbf{\bibinfo{volume}{299}}, \bibinfo{pages}{277--283}
  (\bibinfo{year}{2004}).

\bibitem{jorgensen1973}
\bibinfo{author}{Jorgensen, W.~L.} \& \bibinfo{author}{Salem, L.}
\newblock \emph{\bibinfo{title}{The {{Organic Chemist}}'s {{Book}} of
  {{Orbitals}}}} (\bibinfo{publisher}{{Academic Press}}, \bibinfo{year}{1973}).

\bibitem{wight1974}
\bibinfo{author}{Wight, G.~R.} \& \bibinfo{author}{Brion, C.~E.}
\newblock \bibinfo{title}{K-shell excitation of {{CH4}}, {{NH3}}, {{H2O}},
  {{CH3OH}}, {{CH3OCH3}} and {{CH3NH2}}, by 2.5 {{keV}} electron impact}.
\newblock \emph{\bibinfo{journal}{Journal of Electron Spectroscopy and Related
  Phenomena}} \textbf{\bibinfo{volume}{4}}, \bibinfo{pages}{25}
  (\bibinfo{year}{1974}).

\bibitem{kobayashi1983}
\bibinfo{author}{Kobayashi, K.}
\newblock \bibinfo{title}{Optical spectra and electronic structure of ice}.
\newblock \emph{\bibinfo{journal}{The Journal of Physical Chemistry}}
  \textbf{\bibinfo{volume}{87}}, \bibinfo{pages}{4317--4321}
  (\bibinfo{year}{1983}).

\bibitem{tronc2001}
\bibinfo{author}{Tronc, M.} \& \bibinfo{author}{Azria, R.}
\newblock \bibinfo{title}{Role of morphology of {{D2O}} ice (20
  \textendash{}160 {{K}}) in low energy electron stimulated desorption of
  {{D}}- and {{O}}- anions, and x-ray photon stimulated desorption of {{D}}+
  cations}.
\newblock \emph{\bibinfo{journal}{International Journal of Mass Spectrometry}}
  \bibinfo{pages}{7} (\bibinfo{year}{2001}).

\bibitem{tse2008}
\bibinfo{author}{Tse, J.~S.} \emph{et~al.}
\newblock \bibinfo{title}{X-{{Ray Raman Spectroscopic Study}} of {{Water}} in
  the {{Condensed Phases}}}.
\newblock \emph{\bibinfo{journal}{Physical Review Letters}}
  \textbf{\bibinfo{volume}{100}} (\bibinfo{year}{2008}).

\bibitem{cavalleri2002}
\bibinfo{author}{Cavalleri, M.}, \bibinfo{author}{Ogasawara, H.},
  \bibinfo{author}{Pettersson, L.} \& \bibinfo{author}{Nilsson, A.}
\newblock \bibinfo{title}{The interpretation of {{X}}-ray absorption spectra of
  water and ice}.
\newblock \emph{\bibinfo{journal}{Chemical Physics Letters}}
  \textbf{\bibinfo{volume}{364}}, \bibinfo{pages}{363--370}
  (\bibinfo{year}{2002}).

\bibitem{sellberg2014}
\bibinfo{author}{Sellberg, J.~A.} \emph{et~al.}
\newblock \bibinfo{title}{Comparison of x-ray absorption spectra between water
  and ice: {{New}} ice data with low pre-edge absorption cross-section}.
\newblock \emph{\bibinfo{journal}{The Journal of Chemical Physics}}
  \textbf{\bibinfo{volume}{141}}, \bibinfo{pages}{034507}
  (\bibinfo{year}{2014}).

\bibitem{dartois2015}
\bibinfo{author}{Dartois, E.} \emph{et~al.}
\newblock \bibinfo{title}{Swift heavy ion modifications of astrophysical water
  ice}.
\newblock \emph{\bibinfo{journal}{Nuclear Instruments and Methods in Physics
  Research Section B: Beam Interactions with Materials and Atoms}}
  \textbf{\bibinfo{volume}{365}}, \bibinfo{pages}{472--476}
  (\bibinfo{year}{2015}).

\bibitem{laffon2006}
\bibinfo{author}{Laffon, C.}, \bibinfo{author}{Lacombe, S.},
  \bibinfo{author}{Bournel, F.} \& \bibinfo{author}{Parent, P.}
\newblock \bibinfo{title}{Radiation effects in water ice: {{A}} near-edge x-ray
  absorption fine structure study}.
\newblock \emph{\bibinfo{journal}{The Journal of Chemical Physics}}
  \textbf{\bibinfo{volume}{125}}, \bibinfo{pages}{204714}
  (\bibinfo{year}{2006}).

\bibitem{sieger1998}
\bibinfo{author}{Sieger, M.~T.}, \bibinfo{author}{Simpson, W.~C.} \&
  \bibinfo{author}{Orlando, T.~M.}
\newblock \bibinfo{title}{Production of {{O2}} on icy satellites by electronic
  excitation of low-temperature water ice}.
\newblock \emph{\bibinfo{journal}{Nature}} \textbf{\bibinfo{volume}{394}},
  \bibinfo{pages}{554--556} (\bibinfo{year}{1998}).

\bibitem{reimann1984}
\bibinfo{author}{Reimann, C.} \emph{et~al.}
\newblock \bibinfo{title}{Ion-induced molecular ejection from {{D2O}} ice}.
\newblock \emph{\bibinfo{journal}{Surface Science}}
  \textbf{\bibinfo{volume}{147}}, \bibinfo{pages}{227--240}
  (\bibinfo{year}{1984}).

\bibitem{johnson2011}
\bibinfo{author}{Johnson, R.~E.}
\newblock \bibinfo{title}{Photolysis and radiolysis of water ice}.
\newblock In \emph{\bibinfo{booktitle}{Physics and {{Chemistry}} at {{Low
  Temperatures}}}}, \bibinfo{pages}{297--339} (\bibinfo{year}{2011}).

\bibitem{sieger1997}
\bibinfo{author}{Sieger, M.~T.}, \bibinfo{author}{Simpson, W.~C.} \&
  \bibinfo{author}{Orlando, T.~M.}
\newblock \bibinfo{title}{Electron-stimulated desorption of {{D}}+ from {{D2O}}
  ice: {{Surface}} structure and electronic excitations}.
\newblock \emph{\bibinfo{journal}{Physical Review B}}
  \textbf{\bibinfo{volume}{56}}, \bibinfo{pages}{4925} (\bibinfo{year}{1997}).

\bibitem{ramaker1983b}
\bibinfo{author}{Ramaker, D.~E.}
\newblock \bibinfo{title}{Comparison of photon-stimulated dissociation of
  gas-phase, solid and chemisorbed water}.
\newblock \emph{\bibinfo{journal}{Chemical Physics}}
  \textbf{\bibinfo{volume}{80}}, \bibinfo{pages}{183--202}
  (\bibinfo{year}{1983}).

\bibitem{simpson1998}
\bibinfo{author}{Simpson, W.~C.}, \bibinfo{author}{Orlando, T.~M.},
  \bibinfo{author}{Parenteau, L.}, \bibinfo{author}{Nagesha, K.} \&
  \bibinfo{author}{Sanche, L.}
\newblock \bibinfo{title}{Dissociative electron attachment in nanoscale ice
  films: {{Thickness}} and charge trapping effects}.
\newblock \emph{\bibinfo{journal}{The Journal of Chemical Physics}}
  \textbf{\bibinfo{volume}{108}}, \bibinfo{pages}{5027--5034}
  (\bibinfo{year}{1998}).

\bibitem{lu2001}
\bibinfo{author}{Lu, Q.-B.} \& \bibinfo{author}{Sanche, L.}
\newblock \bibinfo{title}{Enhanced dissociative electron attachment to {{CF}} 2
  {{Cl}} 2 by transfer of electrons in precursors to the solvated state in
  water and ammonia ice}.
\newblock \emph{\bibinfo{journal}{Physical Review B}}
  \textbf{\bibinfo{volume}{63}} (\bibinfo{year}{2001}).

\bibitem{gervais2006}
\bibinfo{author}{Gervais, B.}, \bibinfo{author}{Beuve, M.},
  \bibinfo{author}{Olivera, G.} \& \bibinfo{author}{Galassi, M.}
\newblock \bibinfo{title}{Numerical simulation of multiple ionization and high
  {{LET}} effects in liquid water radiolysis}.
\newblock \emph{\bibinfo{journal}{Radiation Physics and Chemistry}}
  \textbf{\bibinfo{volume}{75}}, \bibinfo{pages}{493--513}
  (\bibinfo{year}{2006}).

\bibitem{pan2004}
\bibinfo{author}{Pan, X.}, \bibinfo{author}{Bass, A.~D.},
  \bibinfo{author}{{Jay-Gerin}, J.-P.} \& \bibinfo{author}{Sanche, L.}
\newblock \bibinfo{title}{A mechanism for the production of hydrogen peroxide
  and the hydroperoxyl radical on icy satellites by low-energy electrons}.
\newblock \emph{\bibinfo{journal}{Icarus}} \textbf{\bibinfo{volume}{172}},
  \bibinfo{pages}{521--525} (\bibinfo{year}{2004}).

\bibitem{petrik2005}
\bibinfo{author}{Petrik, N.~G.} \& \bibinfo{author}{Kimmel, G.~A.}
\newblock \bibinfo{title}{Electron-stimulated sputtering of thin amorphous
  solid water films on {{Pt}}(111)}.
\newblock \emph{\bibinfo{journal}{The Journal of Chemical Physics}}
  \textbf{\bibinfo{volume}{123}}, \bibinfo{pages}{054702}
  (\bibinfo{year}{2005}).

\bibitem{kimmel1994}
\bibinfo{author}{Kimmel, G.~A.}, \bibinfo{author}{Orlando, T.~M.},
  \bibinfo{author}{V{\'e}zina, C.} \& \bibinfo{author}{Sanche, L.}
\newblock \bibinfo{title}{Low-energy electron-stimulated production of
  molecular hydrogen from amorphous water ice}.
\newblock \emph{\bibinfo{journal}{The Journal of Chemical Physics}}
  \textbf{\bibinfo{volume}{101}}, \bibinfo{pages}{3282--3286}
  (\bibinfo{year}{1994}).

\bibitem{kimmel1995}
\bibinfo{author}{Kimmel, G.~A.} \& \bibinfo{author}{Orlando, T.~M.}
\newblock \bibinfo{title}{Low-{{Energy}} (5\textendash{}120 {{eV}})
  {{Electron}}-{{Stimulated Dissociation}} of {{Amorphous D2O Ice}}:
  {{D}}({{2S}}), {{O}}({{3P2}}, 1, 0), and {{O}}({{1D2}}) {{Yields}} and
  {{Velocity Distributions}}}.
\newblock \emph{\bibinfo{journal}{Physical Review Letters}}
  \textbf{\bibinfo{volume}{75}}, \bibinfo{pages}{2606--2609}
  (\bibinfo{year}{1995}).

\bibitem{rocker1990a}
\bibinfo{author}{Rocker, G.} \emph{et~al.}
\newblock \bibinfo{title}{A {{Comparative Study}} of {{Core}}-{{Induced Photon
  Stimulated Desorption}} and {{Electron}}-{{NEXAFS}} of {{Condensed H2O}},
  {{NH3}}, and {{CH4}}}.
\newblock In \bibinfo{editor}{Ertl, G.}, \bibinfo{editor}{Gomer, R.},
  \bibinfo{editor}{Lotsch, H. K.~V.}, \bibinfo{editor}{Betz, G.} \&
  \bibinfo{editor}{Varga, P.} (eds.) \emph{\bibinfo{booktitle}{Desorption
  {{Induced}} by {{Electronic Transitions DIET IV}}}},
  vol.~\bibinfo{volume}{19}, \bibinfo{pages}{261--267}
  (\bibinfo{publisher}{{Springer Berlin Heidelberg}},
  \bibinfo{address}{{Berlin, Heidelberg}}, \bibinfo{year}{1990}).

\bibitem{floyd1972}
\bibinfo{author}{Floyd, G.~R.} \& \bibinfo{author}{Prince, R.~H.}
\newblock \bibinfo{title}{Production of {{Ionized Water Clusters}} by
  {{Electron Bombardment}} of {{Ice}}}.
\newblock \emph{\bibinfo{journal}{Nature Physical Science}}
  \textbf{\bibinfo{volume}{240}}, \bibinfo{pages}{11--12}
  (\bibinfo{year}{1972}).

\bibitem{christiansen1987}
\bibinfo{author}{Christiansen, J.~W.}, \bibinfo{author}{Tsong, I. S.~T.} \&
  \bibinfo{author}{Lin, S.~H.}
\newblock \bibinfo{title}{Ion-induced desorption of
  ({{H}}{\textsubscript{2}}{{O}}){\textsubscript{{\emph{n}}}}{{H}}{\textsuperscript{+}}
  ion clusters}.
\newblock \emph{\bibinfo{journal}{The Journal of Chemical Physics}}
  \textbf{\bibinfo{volume}{86}}, \bibinfo{pages}{4701--4705}
  (\bibinfo{year}{1987}).

\bibitem{martinez2019}
\bibinfo{author}{Martinez, R.} \emph{et~al.}
\newblock \bibinfo{title}{Production of {{Hydronium Ion}} ({{H3O}})+ and
  {{Protonated Water Clusters}} ({{H2O}}){{nH}}+ after {{Energetic Ion
  Bombardment}} of {{Water Ice}} in {{Astrophysical Environments}}}.
\newblock \emph{\bibinfo{journal}{J. Phys. Chem. A}} \bibinfo{pages}{8}
  (\bibinfo{year}{2019}).

\bibitem{jaenicke1986}
\bibinfo{author}{Jaenicke, S.} \emph{et~al.}
\newblock \bibinfo{title}{{{FIELD}}-{{ASSISTED PHOTODESORPTION OF IONS FROM
  METAL AND SEMICONDUCTOR SURFACES}}}.
\newblock \emph{\bibinfo{journal}{Le Journal de Physique Colloques}}
  \textbf{\bibinfo{volume}{47}}, \bibinfo{pages}{C7--343--C7--348}
  (\bibinfo{year}{1986}).

\bibitem{souda2002}
\bibinfo{author}{Souda, R.}
\newblock \bibinfo{title}{Coulomb explosion in electron stimulated desorption
  of positive ions from gas-condensed dielectric surfaces}.
\newblock \emph{\bibinfo{journal}{Surface science}}
  \textbf{\bibinfo{volume}{506}}, \bibinfo{pages}{L275--L281}
  (\bibinfo{year}{2002}).

\bibitem{tachibana2006}
\bibinfo{author}{Tachibana, T.}, \bibinfo{author}{Miura, T.} \&
  \bibinfo{author}{Arakawa, I.}
\newblock \bibinfo{title}{Desorption of water cluster ions from the surface of
  solid rare gases}.
\newblock \emph{\bibinfo{journal}{Low Temperature Physics}}
  \textbf{\bibinfo{volume}{32}}, \bibinfo{pages}{1092--1096}
  (\bibinfo{year}{2006}).

\bibitem{grieves2011}
\bibinfo{author}{Grieves, G.~A.} \& \bibinfo{author}{Orlando, T.~M.}
\newblock \bibinfo{title}{Intermolecular {{Coulomb Decay}} at {{Weakly Coupled
  Heterogeneous Interfaces}}}.
\newblock \emph{\bibinfo{journal}{Physical Review Letters}}
  \textbf{\bibinfo{volume}{107}}, \bibinfo{pages}{016104}
  (\bibinfo{year}{2011}).

\bibitem{baron1976}
\bibinfo{author}{Baron, B.} \& \bibinfo{author}{Williams, F.}
\newblock \bibinfo{title}{X-ray photoelectron spectroscopy of amorphous ice}.
\newblock \emph{\bibinfo{journal}{The Journal of Chemical Physics}}
  \textbf{\bibinfo{volume}{64}}, \bibinfo{pages}{3896--3897}
  (\bibinfo{year}{1976}).

\bibitem{nordlund2004}
\bibinfo{author}{Nordlund, D.} \emph{et~al.}
\newblock \bibinfo{title}{Surface structure of thin ice films}.
\newblock \emph{\bibinfo{journal}{Chemical Physics Letters}}
  \textbf{\bibinfo{volume}{395}}, \bibinfo{pages}{161--165}
  (\bibinfo{year}{2004}).

\bibitem{mase1998}
\bibinfo{author}{Mase, K.} \emph{et~al.}
\newblock \bibinfo{title}{Auger electron photoion coincidence technique
  combined with synchrotron radiation for the study of the ion desorption
  mechanism in the region of resonant transitions of condensed {{H2O}}}.
\newblock \emph{\bibinfo{journal}{The Journal of Chemical Physics}}
  \textbf{\bibinfo{volume}{108}}, \bibinfo{pages}{6550--6553}
  (\bibinfo{year}{1998}).

\bibitem{feulner2000}
\bibinfo{author}{Feulner, P.} \emph{et~al.}
\newblock \bibinfo{title}{Recent progress in the investigation of core
  hole-induced photon stimulated desorption from adsorbates: Excitation
  site-dependent bond breaking, and charge rearrangement}.
\newblock \emph{\bibinfo{journal}{Surface science}}
  \textbf{\bibinfo{volume}{451}}, \bibinfo{pages}{41--52}
  (\bibinfo{year}{2000}).

\bibitem{dujardin1989}
\bibinfo{author}{Dujardin, G.}, \bibinfo{author}{Hellner, L.},
  \bibinfo{author}{Olsson, B.~J.}, \bibinfo{author}{{Besnard-Ramage}, M.~J.} \&
  \bibinfo{author}{Dadouch, A.}
\newblock \bibinfo{title}{Negative-{{Fragment}}-{{Ion Formation}} by {{Photon
  Excitation}} of {{Molecules}} in the {{Vicinity}} of {{Their
  Core}}-{{Ionization}} and {{Direct}}-{{Double}}-{{Ionization Thresholds}}}.
\newblock \emph{\bibinfo{journal}{Physical Review Letters}}
  \textbf{\bibinfo{volume}{62}}, \bibinfo{pages}{745--748}
  (\bibinfo{year}{1989}).

\bibitem{andrade2010}
\bibinfo{author}{Andrade, D. P.~P.}, \bibinfo{author}{Rocco, M. L.~M.} \&
  \bibinfo{author}{{Boechat-Roberty}, H.~M.}
\newblock \bibinfo{title}{X-ray photodesorption from methanol ice: {{Ion}}
  desorption by soft {{X}}-ray}.
\newblock \emph{\bibinfo{journal}{Monthly Notices of the Royal Astronomical
  Society}} \textbf{\bibinfo{volume}{409}}, \bibinfo{pages}{1289--1296}
  (\bibinfo{year}{2010}).

\bibitem{sanche1990}
\bibinfo{author}{Sanche, L.}
\newblock \bibinfo{title}{Low-energy electron scattering from molecules on
  surfaces}.
\newblock \emph{\bibinfo{journal}{Journal of Physics B: Atomic, Molecular and
  Optical Physics}} \textbf{\bibinfo{volume}{23}}, \bibinfo{pages}{1597--1624}
  (\bibinfo{year}{1990}).

\bibitem{bass2003}
\bibinfo{author}{Bass, A.} \& \bibinfo{author}{Sanche, L.}
\newblock \bibinfo{title}{Interactions of {{Low}}-{{Energy Electrons}} with
  {{Atomic}} and {{Molecular Solids}}}.
\newblock In \bibinfo{editor}{Mozumder, A.} \& \bibinfo{editor}{Hatano, Y.}
  (eds.) \emph{\bibinfo{booktitle}{Charged {{Particle}} and {{Photon
  Interactions}} with {{Matter}}}} (\bibinfo{publisher}{{CRC Press}},
  \bibinfo{year}{2003}).

\bibitem{tronc1996}
\bibinfo{author}{Tronc, M.}, \bibinfo{author}{Azria, R.},
  \bibinfo{author}{Le~Coat, Y.} \& \bibinfo{author}{Illenberger, E.}
\newblock \bibinfo{title}{Threefold {{Differential Electron}}-{{Stimulated
  Desorption Yields}} of {{D}}-{{Anions}} from {{Multilayer Films}} of {{D2O}}
  and {{ND3 Condensed}} on {{Platinum}}}.
\newblock \emph{\bibinfo{journal}{The Journal of Physical Chemistry}}
  \textbf{\bibinfo{volume}{100}}, \bibinfo{pages}{14745--14750}
  (\bibinfo{year}{1996}).

\bibitem{simpson1997}
\bibinfo{author}{Simpson, W.~C.}, \bibinfo{author}{Parenteau, L.},
  \bibinfo{author}{Smith, R.~S.}, \bibinfo{author}{Sanche, L.} \&
  \bibinfo{author}{Orlando, T.~M.}
\newblock \bibinfo{title}{Electron-stimulated desorption of {{D}}-({{H}}-) from
  condensed {{D2O}} ({{H2O}}) films}.
\newblock \emph{\bibinfo{journal}{Surface science}}
  \textbf{\bibinfo{volume}{390}}, \bibinfo{pages}{86--91}
  (\bibinfo{year}{1997}).

\bibitem{stolte2003}
\bibinfo{author}{Stolte, W.~C.} \emph{et~al.}
\newblock \bibinfo{title}{Photofragmentation dynamics of core-excited water by
  anion-yield spectroscopy}.
\newblock \emph{\bibinfo{journal}{Physical Review A}}
  \textbf{\bibinfo{volume}{68}} (\bibinfo{year}{2003}).

\bibitem{piancastelli2005}
\bibinfo{author}{Piancastelli, M.~N.} \emph{et~al.}
\newblock \bibinfo{title}{Resonant {{Auger}} decay of above-threshold
  core-excited {{H2O}}}.
\newblock \emph{\bibinfo{journal}{Physical Review A}}
  \textbf{\bibinfo{volume}{71}} (\bibinfo{year}{2005}).

\bibitem{strahlman2016}
\bibinfo{author}{Str{\aa}hlman, C.}, \bibinfo{author}{Kivim{\"a}ki, A.},
  \bibinfo{author}{Richter, R.} \& \bibinfo{author}{Sankari, R.}
\newblock \bibinfo{title}{Negative-{{Ion}}/{{Positive}}-{{Ion Coincidence
  Yields}} of {{Core}}-{{Excited Water}}}.
\newblock \emph{\bibinfo{journal}{The Journal of Physical Chemistry A}}
  \textbf{\bibinfo{volume}{120}}, \bibinfo{pages}{6389--6393}
  (\bibinfo{year}{2016}).

\bibitem{orlando1999}
\bibinfo{author}{Orlando, T.~M.}, \bibinfo{author}{Kimmel, G.~A.} \&
  \bibinfo{author}{Simpson, W.~C.}
\newblock \bibinfo{title}{Quantum-resolved electron stimulated interface
  reactions: {{D2}} formation from {{D2O}} films} \bibinfo{pages}{8}
  (\bibinfo{year}{1999}).

\bibitem{kimmel1996}
\bibinfo{author}{Kimmel, G.~A.} \& \bibinfo{author}{Orlando, T.~M.}
\newblock \bibinfo{title}{Observation of {{Negative Ion Resonances}} in
  {{Amorphous Ice}} via {{Low}}-{{Energy}} (5\textendash{}40 {{eV}})
  {{Electron}}-{{Stimulated Production}} of {{Molecular Hydrogen}}}.
\newblock \emph{\bibinfo{journal}{Physical Review Letters}}
  \textbf{\bibinfo{volume}{77}}, \bibinfo{pages}{3983--3986}
  (\bibinfo{year}{1996}).

\bibitem{pan2005}
\bibinfo{author}{Pan, X.}, \bibinfo{author}{{Abdoul-Carime}, H.},
  \bibinfo{author}{Cloutier, P.}, \bibinfo{author}{Bass, A.~D.} \&
  \bibinfo{author}{Sanche, L.}
\newblock \bibinfo{title}{{{D}}-, {{O}}- and {{OD}}- desorption induced by
  low-energy (0\textendash{{20eV}}) electron impact on amorphous {{D2O}}
  films}.
\newblock \emph{\bibinfo{journal}{Radiation Physics and Chemistry}}
  \textbf{\bibinfo{volume}{72}}, \bibinfo{pages}{193--199}
  (\bibinfo{year}{2005}).

\bibitem{ruhl1991}
\bibinfo{author}{R{\"u}hl, E.} \& \bibinfo{author}{Hitchcock, A.~P.}
\newblock \bibinfo{title}{Oxygen {{K}}-shell excitation spectroscopy of
  hydrogen peroxide}.
\newblock \emph{\bibinfo{journal}{Chemical physics}}
  \textbf{\bibinfo{volume}{154}}, \bibinfo{pages}{323--329}
  (\bibinfo{year}{1991}).

\bibitem{piancastelli1999}
\bibinfo{author}{Piancastelli, M.~N.} \emph{et~al.}
\newblock \bibinfo{title}{Resonant photofragmentation of water at the oxygen
  {{K}} edge by high-resolution ion-yield spectroscopy}.
\newblock \emph{\bibinfo{journal}{Physical Review A}}
  \textbf{\bibinfo{volume}{59}}, \bibinfo{pages}{300} (\bibinfo{year}{1999}).

\bibitem{carlson1983}
\bibinfo{author}{Carlson, T.~A.}
\newblock \bibinfo{title}{The {{Coulomb Explosion}} and {{Recent Methods}} for
  {{Studying Molecular Decomposition}}}.
\newblock In \bibinfo{editor}{Goldanskii, V.~I.} \emph{et~al.} (eds.)
  \emph{\bibinfo{booktitle}{Desorption {{Induced}} by {{Electronic Transitions
  DIET I}}}}, vol.~\bibinfo{volume}{24}, \bibinfo{pages}{169--182}
  (\bibinfo{publisher}{{Springer Berlin Heidelberg}},
  \bibinfo{address}{{Berlin, Heidelberg}}, \bibinfo{year}{1983}).

\end{thebibliography}

\end{document}